\documentclass[pre,aps,showpacs,nofootinbib]{revtex4}
\usepackage{graphicx}
\usepackage{natbib}
\usepackage{bm}
\usepackage{amssymb}
\usepackage{amsmath}
\usepackage{euscript}
\usepackage{esint}
\usepackage{prelim2e}

\begin{document}

\title{Analytical treatment of 2D steady flames anchored
in high-velocity streams}

\author{Kirill A. Kazakov}
\affiliation{%
Department of Theoretical Physics, Physics Faculty, Moscow State
University, 119899, Moscow, Russian Federation}

\begin{abstract}
The problem of burning of high-velocity gas streams in channels is revisited.
Previous treatments of this issue are found to be incomplete. It is shown that despite relative smallness of the transversal gas velocity, it plays crucial role in determining flame structure. In particular, it is necessary in formulating boundary conditions near the flame anchor, and for the proper account of the flame propagation law. Using the on-shell description of steady anchored flames, a consistent solution of the problem is given. Equations for the flame front position and gas-velocity at the front are obtained. It is demonstrated that they reduce to a second-order differential equation for the front position. Numerical solutions of the derived equations are found.

\end{abstract}
\maketitle

\section{Introduction}

The problem of flame propagation in a flow of gaseous mixture with a fixed ignition point is important from both experimental and theoretical points of view. In fact, relative simplicity of practical realization and importance in applications made anchored flames one of the most popular topics in combustion science. Despite these circumstances theoretical description of the subject is far from being complete. It would not even be exaggeration to say that the very mechanism of formation of steady flame configurations is not fully understood. One of the most difficult problems here is the influence of anchoring system on the flame, in particular, the question of its locality. Analytical investigation of this arduous question is so complicated that it is usually not raised at all. Another closely related problem is the identification of mechanisms driving the development of flame disturbances. It is known for a long time that anchored flames may develop bulbous structures \citep{progreport1949}, their stability properties are affected by gravity diversely and deeper than in the case of freely propagating flames \citep{cheng1,cheng2}, which means that the stabilizing mechanisms in the two instances are quite different.

These issues are especially nontrivial in the case of two-dimensional flames. Namely, reduction of dimensionality changes the long-range behavior of the Green functions, leading to enhanced interaction between distant parts of the flame. Specifically, pressure distribution is determined by the Green function of the Laplace operator, whose integral kernel grows logarithmically with distance in two dimensions, and therefore so does the response to a point source. This indicates that the regions with large velocity gradients, in particular, vicinity of the anchor, may have  strong nonlocal impact on the global flame structure.

Of particular interest are the high-velocity streams. Experiments indicate that when the incoming gas velocity significantly exceeds the normal flame burning speed, the flame front assumes highly elongated shape which often can be well approximated by straight lines (V-flames). Even if the front shape is not piecewise linear, simplifications admitted by the high-velocity limit make it accessible for theoretical investigation. An important example is the flame anchored in a high-velocity uniform stream in a channel. The first detailed theoretical account of this case was given by \citep{zel1944} who derived an integral equation for the front position, and obtained its numerical solutions [the main results of this work are reproduced in the book \citep{zeldo1985}]. The problem was considered independently by \citep{scurlock1948}, whose results were subsequently critically reviewed and clarified by \citep{tsien1951}. In the latter work, in particular, the main assumptions employed in the analysis were identified and used to derive an integral equation similar to that of \citep{zel1944}.

All three works deal with the steady case and assume the following:

1) For sufficiently high stream velocity, curvature of the stream lines can be neglected. The gas velocity is thus parallel to the channel walls everywhere. This implies that the gas pressure is constant in every cross section of the fresh and burnt gas regions. Furthermore, neglecting the relatively small constant pressure jump across the front makes pressure constant in every cross section of the channel.

2) The fresh gas velocity is also constant in every cross section, both up- and downstream of the ignition point. Neglecting velocity jump at the front makes the flow field continuous everywhere in the channel.

Using the Bernoulli integral under these assumptions, it is straightforward to show that the pressure is monotonic along the channel, so that it can be taken as an independent coordinate, \citep[see][]{zel1944}. Alternatively, as an independent variable can be taken the fresh gas velocity which is also monotonic along the channel, \citep[see][]{tsien1951}. In either way, the use of mass conservation gives an integral equation for one of the flow variables.

It should be mentioned that no attempt was made in the cited papers to justify the approximations 1), 2) any more rigorously than outlined above. Although these approximations look naturally, exclusion of the transversal velocity component from the list of dynamical variables represents a quite nontrivial step. Indeed, replacing the system of two Euler equations by the single Bernoulli integral means that the two velocity components are completely decoupled from each other. Considered on its own,  this reduction of the system is legitimate under the assumption of high stream velocity. But the flow equations themselves do not constitute the complete system of equations governing flame propagation. They must be supplemented by the evolution equation and the jump conditions at the flame front, as well as boundary conditions at the channel walls. But the two velocity components are strongly coupled by the evolution equation: relatively small variations of the transversal component give rise to large variations of the longitudinal component [Cf. Eq.~(\ref{evolutiongen1}) below]. At the same time, exclusion of the transversal component from the consideration based on the assumptions 1), 2) makes this essential equation obsolete. Thus, in order that the evolution equation take its proper place in the analysis of flame propagation, it is necessary to bring the transversal velocity component back into consideration. This requires in turn restoration of the remaining flow equation and the corresponding jump condition.

An important step in validating the model described above was made by \citep{cherny1954} who showed that the equations derived by Zel'dovich and Tsien are asymptotically exact. More precisely, he formulated the way the high-velocity limit should be taken in the complete system of governing equations, and demonstrated that the resulting system reduces to an integral equation which is equivalent to the equations derived by these authors. One of the purposes of the present paper will be to show that Cherny's consideration is not complete. Namely, it omits one of the boundary conditions to be satisfied by the flow velocity in the bulk. It turns out that enforcement of the missing condition makes the integral equation trivial, in the sense that the only its solution satisfying all boundary conditions is the rectilinear front configuration with constant upstream gas velocity.

Evidently, this fact entails two conclusions. First, the assumptions 1), 2) mentioned above oversimplify the problem, and cannot be used to describe nontrivial flame configurations. Second, the limiting procedure proposed by \citep{cherny1954} is also inadequate. The main purpose of the present paper is to derive the correct equation describing flames in high-velocity streams, and to find its nontrivial solutions.
For this purpose, the on-shell description of steady flames, developed in \citep{kazakov1,kazakov2}, will be used. Its main advantage is that it describes flames in a closed form, {\it i.e.,} in a form involving only quantities defined on the flame front. There is no need to solve the flow equations in the bulk explicitly. In particular, it allows one to avoid artificial assumptions about the bulk flow, such as the scaling laws for the gas-velocity and pressure in the high-velocity limit, which are adopted in one way or another by the conventional approach. This description was initially given for freely propagating flames, but it admits simple and natural extension to anchored flames. This generalization is obtained in \citep{jerk3}.

The paper is organized as follows. The Zel'dovich-Scurlock-Tsien approach is critically reviewed in Sec.~\ref{critiques} which starts with a brief account of Cherny's formulation of the high-velocity limit. It is shown, in particular, that this formulation reproduces the assumptions 1),2) of Zel'dovich-Scurlock-Tsien approach, and that the only solution of the main equation, satisfying boundary conditions, is the trivial solution. The reasons underlying this result are identified in Secs.~\ref{boundarycond}, \ref{roleeq}. We then go to the on-shell description in Sec.~\ref{onshell}, summarizing the main equations derived in \citep{kazakov1,kazakov2}, and their extension to anchored flames. The reader is referred to \citep{jerk3} for more details concerning inclusion of the anchoring system and its analytical description. Sections \ref{rolep} -- \ref{evolutionequation} are counterparts of Secs.~\ref{chernyf} -- \ref{roleeq}. They discuss the role of pressure, and indicate the place the boundary conditions and evolution equation take in our approach. Solution of the on-shell equations is obtained in Sec.~\ref{solution}. In Sec.~\ref{largesl}, a large-slope expansion of these equations is derived, which extends to curved flames the corresponding expansion constructed in \citep{jerk3} for V-flames. Using this expansion, it is shown in Sec.~\ref{reduction1} that the main integro-differential equation for the complex velocity reduces to ordinary differential equations. Together with the evolution equation, these equations can be partially integrated and further reduced to a single second-order differential equation. This is done in Sec.~\ref{reduction} where numerical solutions of the derived equations are also found. Section \ref{conclusions} summarizes the results of the work. The paper has an appendix which describes in detail transition to the case of vanishingly small anchor dimensions within the large-slope expansion.

\section{Critiques of Zel'dovich-Scurlock-Tsien approach}\label{critiques}

\subsection{Cherny's formulation of the high-velocity limit}\label{chernyf}

\begin{figure}
\centering
\includegraphics[width=.4\textwidth]{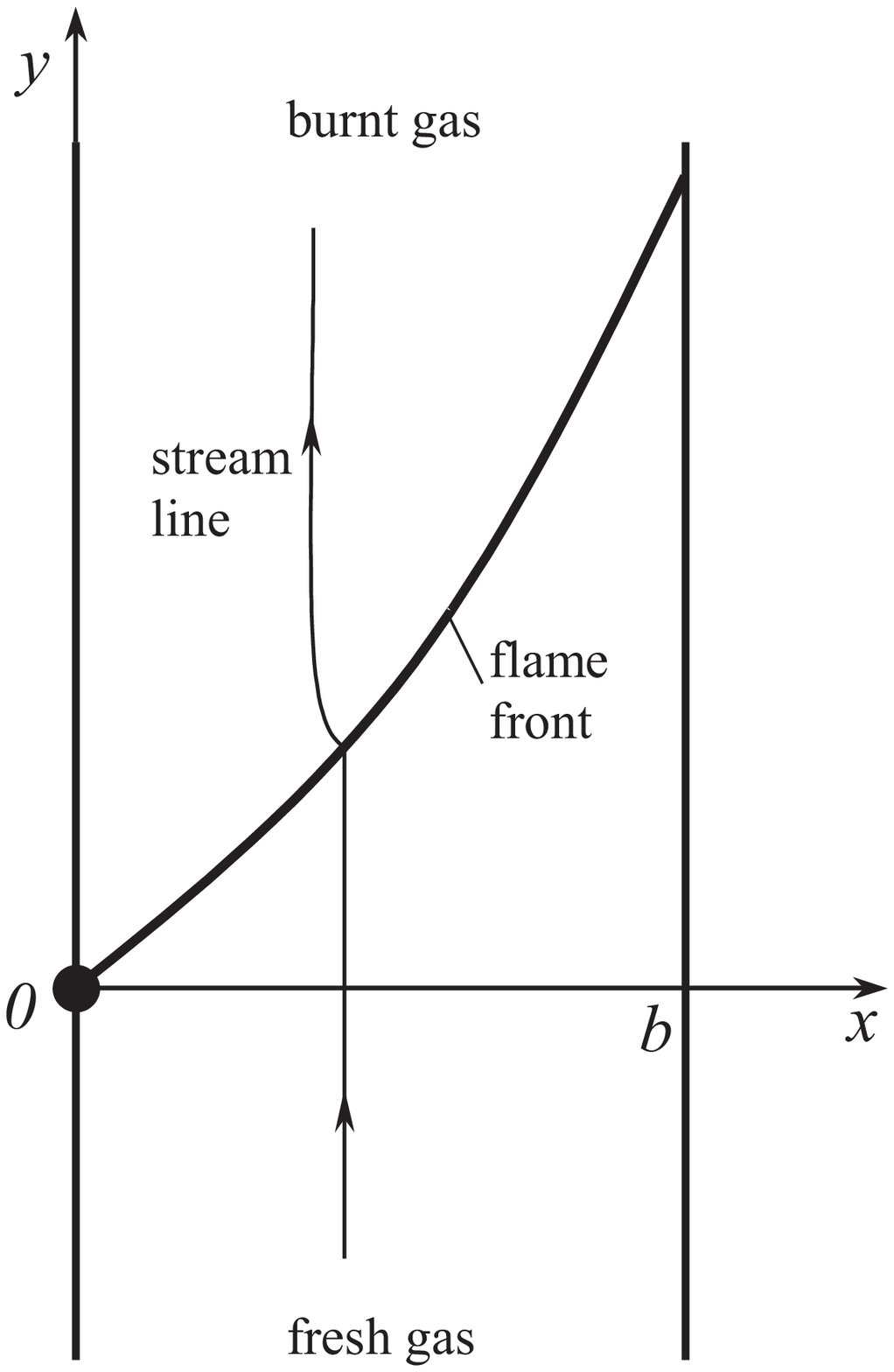}
\caption{}\label{fig1}
\end{figure}

We start with brief recalling the results of \citep{cherny1954} using notation appropriate for the present paper. These results provide a rigorous base for Zel'dovich-Scurlock-Tsien approach. Consider a steady two-dimensional combustible ideal gas stream in a channel with plane-parallel walls, ignited at a fixed point in the middle of the channel (see Fig.~\ref{fig1} where only the right half of the channel is shown). This point (the flame anchor) will be chosen as the origin of Cartesian system of coordinates $\bm{r} = (x,y),$ with $y$-axis along the wall, and the initially uniform fresh gas at $y=-\infty.$ The channel half-width will be taken as a unit of length, while the gas velocity $\bm{v} = (w,u)$ will be measured in units of the normal flame velocity relative to the fresh gas. Then the flow variables obey the following equations in the bulk
\begin{eqnarray}\label{flow1}
\frac{\partial w}{\partial x} + \frac{\partial u}{\partial y} &=& 0\,,
\\ w\frac{\partial w}{\partial x} + u\frac{\partial w}{\partial y} &=& - \frac{1}{\rho}\frac{\partial p}{\partial x} \,, \label{flow2}
\\ w\frac{\partial u}{\partial x} + u\frac{\partial u}{\partial y} &=& - \frac{1}{\rho}\frac{\partial p}{\partial y} \,, \label{flow3}
\end{eqnarray}
\noindent where $p,\rho$ are respectively the gas pressure and density. Under the assumption that the flow is essentially subsonic, the gas can be considered incompressible. Taking the fresh gas density as a density unit, that of the burnt gas will be $1/\theta,$ where $\theta>1$ is the gas expansion coefficient.
The flame configuration is assumed to be symmetric with respect to the $y$-axis. More precisely,
\begin{eqnarray}\label{antisym}
f(x) = f(-x)\,, \qquad w(x,y) = - w(-x,y)\,, \qquad u(x,y)
= u(-x,y)\,.
\end{eqnarray}
\noindent In particular, the transversal velocity component vanishes at the symmetry axis, $w(0,y) = 0,$ so that this formulation applies also to the case of a flame anchored at the channel wall, the point of view taken up in \citep{zel1944}.

Cherny formulated the high-velocity limit as follows. Denote the fresh gas velocity far from the flame front by $U,$ and introduce new (designated with a tilde) coordinates and flow variables according to
\begin{eqnarray}\label{newvar}
\tilde{x} = x\,, \quad \tilde{y} = y/U\,, \quad \tilde{u} = u/U\,, \quad
\tilde{w} = w\,, \quad \tilde{p} = p/U^2\,.
\end{eqnarray}
\noindent After that, switch to the new independent variables $(\tilde{y},\psi),$ where $\psi$  is the stream function defined by
\begin{eqnarray}\label{psivar}
\rho \tilde{u} = \frac{\partial \psi}{\partial \tilde{x}}\,, \quad \rho \tilde{w} = - \frac{\partial \psi}{\partial \tilde{y}}\,.
\end{eqnarray}
\noindent
Then the limit $U \to \infty$ is to be taken under the assumption that the quantities $\tilde{w},\tilde{u},\tilde{p}$ as well as their derivatives with respect to $\tilde{y},\psi$ remain bounded. Equations~(\ref{flow2}), (\ref{flow3}) (\ref{psivar}) thus take the form
\begin{eqnarray}\label{floweq}
\frac{\partial \tilde{p}}{\partial\psi} = 0\,, \quad \tilde{u}\frac{\partial \tilde{u}}{\partial \tilde{y}} + \frac{1}{\rho}\frac{\partial \tilde{p}}{\partial \tilde{y}} = 0\,, \quad \frac{\partial \tilde{x}}{\partial\psi} = \frac{1}{\rho \tilde{u}}\,, \quad \frac{\partial \tilde{x}}{\partial\tilde{y}} = \frac{\tilde{w}}{\tilde{u}}\,.
\end{eqnarray}
\noindent The first two equations give
\begin{eqnarray}\label{floweqint}
\tilde{p} = \tilde{p}(\tilde{y})\,, \quad \frac{\tilde{u}^2}{2} + \frac{\tilde{p}}{\rho} = i(\psi)\,,
\end{eqnarray}
\noindent where $i(\psi)$ is an arbitrary function. There are also jump conditions to be satisfied across the flame front, which in the limit $U \to \infty$ read
\begin{eqnarray}\label{boundclim}
\tilde{u}_+ = \tilde{u}_-\,, \quad \tilde{p}_+ = \tilde{p}_-\,, \quad \psi_+ = \psi_-\,,
\end{eqnarray}
\noindent where the minus (plus) subscript denotes restriction to the flame front of the flow function defined upstream (downstream). Noting that $i(\psi) = {\rm const}$ upstream (since the incoming flow is uniform), we see that equations (\ref{floweqint}) together with the first two equations (\ref{boundclim}) exactly reproduce the assumptions 1) and 2) stated in the introduction. Next, integrating the third of equations (\ref{floweq}), and combining its solution with the relations~(\ref{floweqint}) Cherny arrives at the following integral equation for the function $\tilde{y}(\tilde{u}_-)$
\begin{eqnarray}\label{zteq}
(\theta - 1)\int\limits_{1}^{u_1}\frac{u\tilde{y}(u)du}{\sqrt{\theta u^2_1 - (\theta - 1)u^2}} = \frac{(u_1 - 1)^2}{2}\,, \quad u_1\equiv \tilde{u}_-\,,
\end{eqnarray}
\noindent which is equivalent to the equations derived by Zel'dovich and Tsien. This completes the proof that the assumptions 1), 2) of Zel'dovich-Scurlock-Tsien approach are equivalent to the limiting procedure formulated by Cherny.

But this is not the end of the story. It turns out that the consideration of \citep{cherny1954} is not complete, in that it does not take into account boundary conditions for the gas flow. More precisely, these conditions are applied only at the end-points of the flame front, but not in the bulk. In terms of the stream function, the condition that the transversal component, $w,$ of gas velocity vanish at the walls and at the symmetry axis reads
\begin{eqnarray}\label{bcpsi1}
\psi = 0 \quad {\rm for} \quad \tilde{x} = 0, \\
\psi = 1 \quad {\rm for} \quad \tilde{x} = 1, \label{bcpsi2}
\end{eqnarray}
\noindent as is seen from equations~(\ref{psivar}) written in the form
$d\psi = -\rho \tilde{w}d\tilde{y} + \rho \tilde{u}d\tilde{x}.$
Consider the flow upstream. Since $i(\psi) = {\rm const}$ there, Eqs.~(\ref{floweqint}) tell us that $u$ is independent of $\psi,$ so that integration of the third of Eqs.~(\ref{floweq}) yields
$$\tilde{x} = \frac{\psi}{\tilde{u}(\tilde{y})} + X(\tilde{y})\,,$$ where the function $X(\tilde{y})$ is to be determined from the boundary conditions (\ref{bcpsi1}), (\ref{bcpsi2}). Substitution gives $$X(\tilde{y}) = 0\,, \quad \tilde{u}(\tilde{y}) = 1\,.$$ Therefore, the flow turns out to be uniform in the whole region upstream of the flame front. Then Eqs.~(\ref{floweqint}), (\ref{boundclim}) show that so is the flow downstream. Thus, $u_1=1$ is the only solution of Eq.~(\ref{zteq}) consistent with the boundary conditions at the channel walls. In other words, the limiting procedure proposed by Cherny, and hence the equivalent assumptions 1), 2) cannot be used to describe nontrivial flame configurations.

Technically, this means that the scalings (\ref{newvar}) are inappropriate. For instance, the scaling of pressure only seems natural. It is suggested by the Bernoulli integral
$$\frac{p}{\rho} + \frac{1}{2}(u^2 + w^2) = {\rm const}$$ which after neglecting $w$ in comparison with large $u$ would give $p \sim U^2.$ However, it is variable part of pressure that only matters. Setting $u = U + \hat{u},$ and absorbing $U^2/2$ in the ``{\rm const}'' on the right gives
$$\frac{p}{\rho} + U\hat{u} + \frac{\hat{u}^2}{2} = {\rm const}\,.$$ Now, $p\sim U,$ if $\hat{u}$ is assumed bounded, or something else if $\hat{u}$ behaves differently.

\subsection{Boundary conditions}\label{boundarycond}

\begin{figure}
\centering
\includegraphics[width=.5\textwidth]{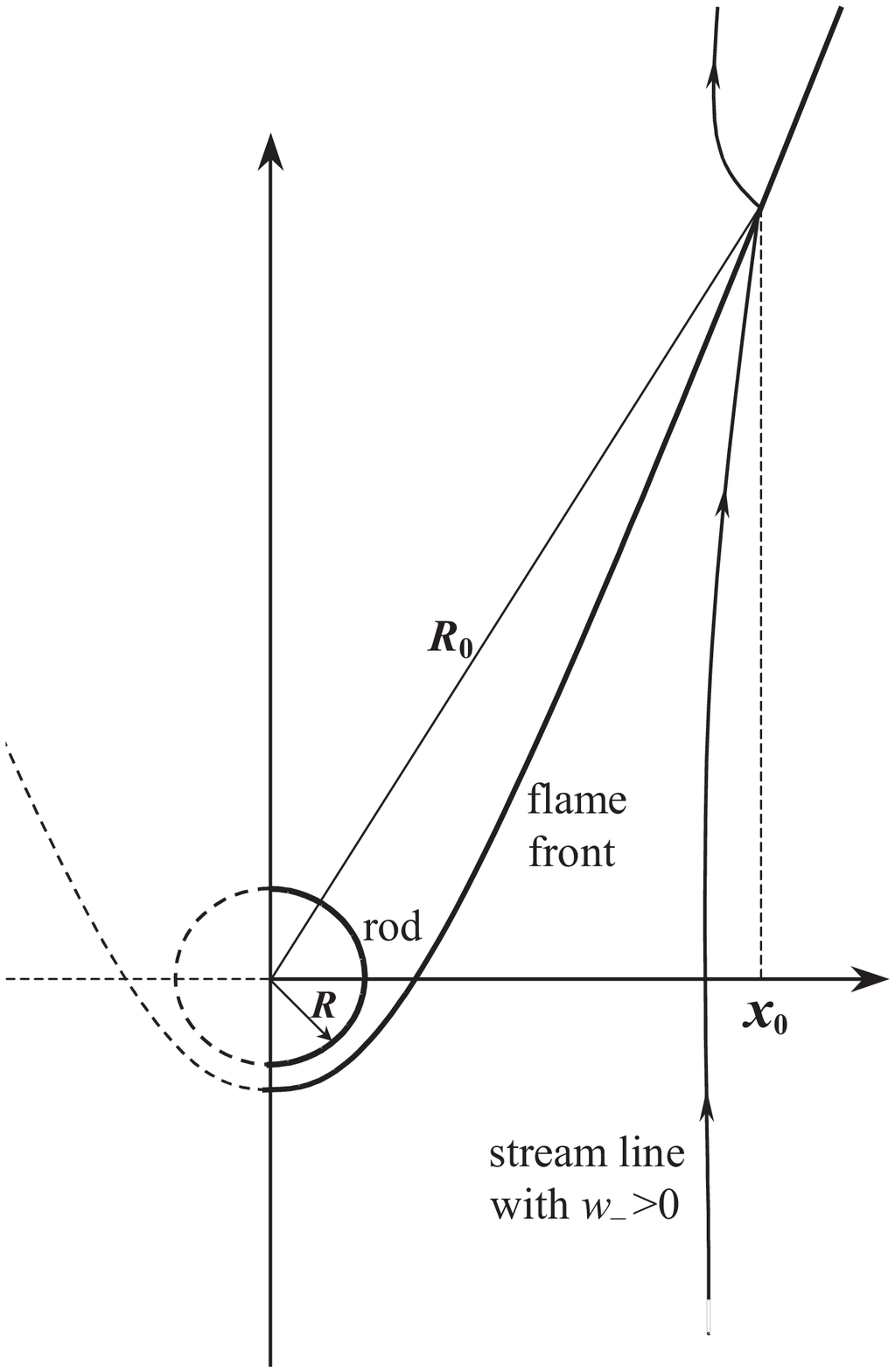}
\caption{}\label{fig2}
\end{figure}

Although not always stated explicitly, impermeability of the channel walls (and of the symmetry axis) is assumed in the Zel'dovich-Scurlock-Tsien approach, but is not used. As we saw in the preceding section, enforcing this condition in the Cherny's formulation makes the flow trivial, leaving the piecewise linear front configuration with constant gas velocities up- and downstream as the only possibility. In this formulation, impermeability of the walls and the symmetry axis is expressed in the form (\ref{bcpsi1}), (\ref{bcpsi2}), but these conditions are used only at the end-points of the flame front, and at the origin. But in fact, things must be just the opposite: transversal gas velocity must vanish at the channel walls and the symmetry axis everywhere except a small vicinity of the flame anchor. Indeed, suppose for simplicity that the flame is anchored by a cylindrical rod with circular cross-section (as is often the case in practice, see Fig.~\ref{fig2}). Then the gas flow is significantly disturbed in a vicinity of the rod: The slowdown of gas elements heading the rod leads to appearance of a nonzero transversal velocity component. This happens no matter how small the rod radius is. If one follows the trajectory of a fresh-gas element moving near the symmetry axis, its transversal  velocity rapidly changes near the rod from zero to some finite value at the flame front. This means that in the limit of vanishing anchor dimensions ({\it i.e.,} in the Zel'dovich-Scurlock-Tsien setting), the gas flow is singular at the point of its location.

To be more specific, the presence of the rod in a high-velocity stream can be described mathematically by superimposing the incoming flow velocity $\bm{v}^0 = (0,U)$ with the velocity field $\bm{v}^d$ of a dipole located at the origin
\begin{eqnarray}\label{dipole}
\bm{v}^d = \frac{UR^2}{r^4} (-2xy, x^2 - y^2)\,,
\end{eqnarray}
\noindent where $R \ll 1$ is the rod radius. Indeed, if the rod is located downstream as is shown in Fig.~\ref{fig2}, the flame front bends round the rod, and for large $U,$ gets close to its surface, so that the normals to the front and the rod surface coincide. To the leading order, the normal gas-velocity is negligible in comparison with $U.$ Hence, velocity of the fresh gas near the rod must satisfy
\begin{eqnarray}\label{dipoleident}
(\bm{v},\bm{\nu}) = 0\,,
\end{eqnarray}
\noindent
where $\bm{\nu}$ is the normal to the rod. The velocity field $\bm{v}^0 + \bm{v}^d$ does satisfy this condition by virtue of the identity $(\bm{v}^0 + \bm{v}^d,\bm{r})|_{r=R} = 0,$ because $\bm{r}$ is normal to the rod. On the other hand, if the rod is located upstream ({\it i.e.,} flame is stabilized in the wake of the rod), Eq.~(\ref{dipoleident}) is the true boundary condition obeyed by $\bm{v}^0 + \bm{v}^d$ identically.

The transversal velocity component induced by the rod is large for $r\sim R,$ and rapidly decreases with distance. At distances $r$ such that  $R\ll r \ll 1,$ the inner solution describing the flow near the rod is to be matched with the large-scale outer solution we are interested in. In particular, matching at the flame front assigns $w_-$ a definite value, say~$w_0.$ Since the transversal velocity in the outer flow is of the order of unity, so is $w_0.$ Let us denote by $R_0,$ $R\ll R_0 \ll 1,$ the characteristic distance where the two solutions are matched near the flame front. We thus have (considering the right branch of the front, $x>0$)
\begin{eqnarray}\label{bcpsir}
w_-|_{r\sim R_0} = w_0\,,
\end{eqnarray}
\noindent where $w_0$ is a positive or negative number, while  \begin{eqnarray}\label{bcpsiry}
w(0,y)=0\,, \quad y<-R.
\end{eqnarray}
Finally, let us determine the scaling of $R_0$ with $U.$ For $r\sim R_0,$ $y=f(x),$ one has $y \sim R_0,$ $x\sim R,$ and hence $$w^d_-|_{r\sim R_0}\sim \frac{UR^2}{R^4_0}RR_0 = \frac{UR^3}{R^3_0}\,.$$ Therefore, $w_0 \sim 1$ implies $R_0\sim U^{1/3}R.$
Taking this into account, we find also
$$u^d_-|_{r\sim R_0}\sim \frac{UR^2}{R^4_0}R^2_0 = U^{1/3}\,.$$
\noindent Neglecting $U^{1/3}$ in comparison with $U,$ we conclude that to the leading order, the longitudinal velocity satisfies
\begin{eqnarray}\label{bcpsiru}
u_-|_{r\sim R_0} = U\,.
\end{eqnarray}
\noindent
Needless to say that the boundary behavior outlined above cannot be described within the Zel'dovich-Scurlock-Tsien approach in which transversal velocity component is completely excluded from consideration.

We have considered the simplest case of a circular rod. In the general case, specific form of the local flow is of course different, but the hierarchy of length scales $R\ll R_0 \ll 1,$ as well as relations (\ref{bcpsir}), (\ref{bcpsiru}) remain the same. These relations will be invoked in Sec.~\ref{strategy}.

\subsection{The role of evolution equation}\label{roleeq}

The local propagation law of a flame is determined by the so-called evolution equation which gives the normal fresh gas velocity, $v^n_-,$ as a function of the flame front curvature and space-time derivatives of the gas velocity at the front. For zero-thickness flames, it states that the normal flame velocity is simply equal to the velocity of planar flame. Namely, if the flame front  position is represented by the curve $y=f(x),$ the evolution equation reads $v^n_- = 1,$ or
\begin{eqnarray}\label{evolutiongen}
u_- - f'w_- = N\,,
\end{eqnarray}
\noindent where the prime denotes $x$-differentiation, and $N = \sqrt{1 + f'^2}.$
Curiously, this characteristic property which plays fundamental role in the whole theory of flame propagation is not invoked in the derivation of the  Zel'dovich-Tsien equation. The reason for this is the already mentioned disregard of the transversal velocity component. The point is that the transversal and longitudinal velocity components are strongly coupled by the evolution equation. Indeed, $u_-$ and $f'$ are both $O(U),$ so that the two terms on the left-hand side of Eq.~(\ref{evolutiongen}) are of the same order of magnitude. Hence, exclusion of $w$ from the list of dynamical variables makes this equation obsolete. In other words, any surface of discontinuity with arbitrary propagation law would satisfy Eq.~(\ref{zteq}), provided that its normal velocity is small compared to $U.$

The only place where Eq.~(\ref{evolutiongen}) finds application in the Zel'dovich-Scurlock-Tsien approach is the relation between the flame front length and  the incoming stream velocity
\begin{eqnarray}\label{lu}
U = f(1)\,,
\end{eqnarray}
\noindent which follows from the mass conservation condition
$$U = \int\limits_{0}^{1}dx \sqrt{1 + f'^2}\, v^n_- \approx \int\limits_{0}^{1}dx f' = f(1) - f(0)\,.$$ Of course, this relation is independent of the particular dynamical model employed, and is valid in the high-velocity limit whatever the structure of the incoming flow. In the Cherny's formulation, Eq.~(\ref{evolutiongen}) takes the form $d\psi_-/d\tilde{y} = 1\,,$ and after integration under conditions (\ref{bcpsi1}), (\ref{bcpsi2}) yields Eq.~(\ref{lu}). These equations are not used in the derivation of Eq.~(\ref{zteq}).

\section{On-shell description of flame propagation}\label{onshell}

As shown in \citep{kazakov1,kazakov2}, the system of bulk flow equations and jump conditions at the flame front can be reduced to a single complex integro-differential equation relating values of the flow variables at the flame front (their {\it on-shell} values), so that explicit solving of the bulk equations (which is the most difficult part of consideration) turns out to be unnecessary. This equation reads
\begin{eqnarray}\label{generalc1st}&&
2\left(\omega_-\right)' + \left(1 +
i\hat{\EuScript{H}}\right)\left\{[\omega]' -
\frac{Nv^n_+\sigma_+\omega_+}{v^2_+} \right\} = 0\,.
\end{eqnarray}
\noindent Here, $\omega = u + iw$ is the complex velocity, $[\omega] = \omega_+ - \omega_-$ its jump across the front, $v^n_+$ is the normal velocity of burnt gas at the front, $\sigma_+$ is the on-shell value of vorticity produced by the curved flame, and the operator $\hat{\EuScript{H}}$ is defined on $2$-periodic functions  by
\begin{eqnarray}\label{hcurvedf}
\left(\hat{\EuScript{H}}a\right)(x) = \frac{1 + i
f'(x)}{2}~\fint\limits_{-1}^{+1}
d\eta~a(\eta)\cot\left\{\frac{\pi}{2}(\eta - x +
i[f(\eta) - f(x)])\right\}\,,
\end{eqnarray}
\noindent the slash denoting the principal value of integral. It satisfies the important identity
\begin{eqnarray}\label{hident}
\hat{\EuScript{H}}^2 = - 1\,.
\end{eqnarray}
\noindent
The functions $\sigma_+,v^n_+$ as well as the velocity jumps at the front, entering equation (\ref{generalc1st}), are all known functionals of on-shell fresh gas velocity \citep[see, e.g.,][]{matalon,pelce}. For zero-thickness flames,
\begin{eqnarray}\label{jumps}
\bar{v}^n_+ &=& \theta\,, \quad [u] = \frac{\theta - 1}{N}\ ,
\quad [w] = - f'\frac{\theta - 1}{N}\,, \\
\sigma_+ &=& - \frac{\theta - 1}{2\theta N}(u^2_- + w^2_-)'\,.\label{vorticity}
\end{eqnarray}
\noindent Together with the evolution equation
(\ref{evolutiongen}), Eq.~(\ref{generalc1st}) constitutes a closed system of three equations for the three unknown functions $w_-(x), u_-(x),$ and $f(x).$

Equation (\ref{generalc1st}) describes freely propagating flames, but can be easily modified to take into account the presence of the rod. This generalization is given in \citep{jerk3}, and reads
\begin{eqnarray}\label{generalc1str}&&
2\left(\omega_-\right)' + \left(1 +
i\hat{\EuScript{H}}\right)\left\{[\omega]' -
\frac{Nv^n_+\sigma_+\omega_+}{v^2_+} \right\} = 2\left(\omega^d_-\right)'\,,
\end{eqnarray}
\noindent where $\omega^d$ is the complex velocity of the dipole (\ref{dipole}). Since this field satisfies the symmetry relations (\ref{antisym}), the boundary condition (\ref{bcpsiry}) is still met. We assume for definiteness that the rod is located in the downstream region (as is usually the case in practice; this assumption is inconsequential, Cf. footnote~\ref{footn1}). Then  $\omega_-$ and $\omega^d_-$ satisfy
\begin{eqnarray}\label{chup}
\left(1 - i \hat{\EuScript{H}}\right)\left(\omega_-\right)' &=& 0\,.\\ \label{chupd}
\left(1 - i\hat{\EuScript{H}}\right)\left(\omega^d_-\right)' &=& 0\,,
\end{eqnarray}
\noindent which express analyticity and boundedness of the functions $\omega(z),$ $\omega^d(z),$ $z=x+iy,$ in the upstream region. Equations (\ref{chup}), (\ref{chupd}) are consistent with Eq.~(\ref{generalc1str}) by virtue of the identity (\ref{hident}).

We will now consider the problem of anchored flame propagation within the on-shell description in the light of the issues discussed in Sec.~\ref{critiques}.

\subsection{The role of gas pressure}\label{rolep}

It was already mentioned in the Introduction that specifics of the two-dimensional case make the issue of flame flow nonlocality especially nontrivial. The region near the rod is characterized by large velocity gradients, and therefore affects significantly the pressure field far apart from the rod. Indeed, pressure is determined by the Poisson equation $$\Delta p = -\rho \left(\nabla(\bm{v}\nabla)\bm{v}\right)\,,$$ while the Green function of the Laplacian is proportional to $\ln r.$  The flow near the anchor thus can be expected to have a strong nonlocal influence on the flame structure. It is one of the advantages of the on-shell formulation that it reveals the completely subordinate role pressure plays in determining the front structure. In fact, the variable $p$ does not appear in Eq.~(\ref{generalc1st}). In particular, no assumption such as that contained in the point 1) of the Zel'dovich-Scurlock-Tsien approach, or the scaling prescription (\ref{newvar}) of the Cherny's limiting procedure, is needed in the on-shell description. Also, exclusion of pressure from consideration weakens the above argument concerning nonlocality of the anchor impact.

\subsection{Strategy of solving Eq.~(\ref{generalc1str}). Boundary conditions}\label{strategy}

We are concerned with the situation where the rod radius, $R,$ is much smaller than the channel width, and interested in the large-scale front structure, {\it i.e.,} its structure at distances large compared to $R$ (the outer solution). We observe, first of all, that the presence of the rod is described simply by adding the dipole field to Eq.~(\ref{generalc1st}), which is noticeable only in a small vicinity of the origin. This fact indicates that despite the pressure argument given in the preceding section, the detailed flow structure near the rod may be actually unimportant. This would mean that the rod can be considered point-like, thus greatly simplifying account of its influence on the global flame structure. Considerations of the subsequent sections will show that this is indeed so in the high-velocity limit. Namely, it turns out that Eq.~(\ref{generalc1st}) reduces in this limit to an ordinary differential equation. This naturally opens the following way of solving Eq.~(\ref{generalc1str}): we consider this equation at $x$'s such that $r\gtrsim R_0,$ and extract the leading terms of the high-velocity expansion. Detailed analysis reveals that for such $x$'s, the dipole contribution is negligible, so that Eq.~(\ref{generalc1str}) reduces to Eq.~(\ref{generalc1st}). Exclusion of the region near the rod implies that this equation must be supplemented by an auxiliary condition at a point $x_0$ corresponding to the matching region $r\sim R_0$ (see Fig.~\ref{fig2}). Since Eq.~(\ref{generalc1st}) is complex, and involves only quantities defined at the flame front, it requires two real auxiliary conditions expressed in terms of on-shell quantities. As such, we take the relations (\ref{bcpsir}), (\ref{bcpsiru}) in which the numbers $w_0,U$ play the role of parameters specifying given large-scale solution. In the limit $R\to 0,$ these relations take the form
\begin{eqnarray}\label{bcpsir1}
w_-(0^+) = w_0\,,\\
\label{bcpsiru1}
u_-(0^+) = U\,,
\end{eqnarray}
\noindent with the understanding that the functions $w_-(x)\equiv w(x,f(x)),$ $u_-(x)\equiv u(x,f(x))$ are solutions of Eq.~(\ref{generalc1st}) considered on the semi-open interval $x \in (0,1],$ and the quantities $w_-(0^+),$ $u_-(0^+)$ are their right limiting values for $x \to 0.$ Furthermore, taking the limit $R\to 0$ leads obviously to the following condition for the function $f(x)$
\begin{eqnarray}\label{bcpsirf}
f(0) = 0\,.
\end{eqnarray}
\noindent
Finally, there is also the usual condition of vanishing of $w$ at the wall $x = +1,$
\begin{eqnarray}\label{bcpsirw1}
w_-(+1) = 0\,.
\end{eqnarray}
\noindent If a solution of Eqs.~(\ref{evolutiongen}), (\ref{generalc1st}) satisfying the above conditions is found for $x \in (0,1],$ then the rule (\ref{antisym}) gives it on the semi-open interval $x \in [-1,0)$ as
\begin{eqnarray}\label{antisym1}
f(x) = f(-x)\,, \qquad w_-(x) = - w_-(-x)\,, \qquad u_-(x)
= u_-(-x)\,,
\end{eqnarray}
\noindent in particular, $w_-(x)$ satisfies $w_-(0^-) = -w_0,$ so that this function has a discontinuity at $x=0.$

In connection with the above procedure of finding the large-scale solution, the following circumstance should be emphasized.  We are able to formulate this procedure in a closed form, without the need to construct the local solution near the rod explicitly, just because Eq.~(\ref{generalc1st}) relates only functions defined at the flame front. This would not be possible in the conventional approach based on explicit solving the bulk equations. Indeed, as we saw in Sec.~\ref{boundarycond},  the transversal velocity of gas elements moving near the rod rapidly changes from zero to some finite value at the flame front. Thus, if we were to construct a bulk outer solution satisfying the matching and boundary conditions (\ref{bcpsir}), (\ref{bcpsiry}), this solution ought to describe the rapid change of $w$ from zero to $w_0 \sim 1$ over a distance $\sim R_0.$ One might try to avoid this complication by shifting the matching region farther from the rod, but then the dipole field would be completely neglected. In effect, the anchor disappears from the description, and the solution becomes trivial. This is exactly what happens in the
Zel'dovich-Scurlock-Tsien approach. In our approach, on the contrary, only longitudinal velocity induced by the dipole is negligible in the matching region, but not the transversal.

At last, it is worth noting that the parameters $w_0, U$ specifying the large-scale solution are independent of each other. This fact becomes evident when we note that instead of the rod with circular cross-section we might use a more complicated cylindrical shape. Then for the same value of $U,$ matching of the inner and outer solutions would give a different value for $w_0.$ We conclude that the large-scale solutions form a two-parameter family. This is in contrast with the
Zel'dovich-Scurlock-Tsien approach where the single parameter $U$ completely determines the solution.

\subsection{Evolution equation}\label{evolutionequation}

Having formulated the outer problem in a closed form, we may now note that the experimentally observed anchored flames usually have fairly smooth, highly elongated front configurations. Hence, neglecting the small regions near the rod and the channel walls where finite-thickness effects are only important, we can use the evolution equation in the simplest form (\ref{evolutiongen}) applicable to zero-thickness flames. Restricting ourselves again to the right half of the channel where the front slope is positive, and taking into account that $f' = O(U)$ [Cf. Eq.~(\ref{lu})]  this equation can be rewritten as
\begin{eqnarray}\label{evolutiongen1}
u_- = f'(w_- + 1)\,, \quad x>0\,,
\end{eqnarray}
\noindent the correction term being of the relative order $O(1/U^2).$

It should be emphasized that neglecting the regions near the rod and the channel walls makes boundary conditions for the function $f'(x)$ itself unnecessary. Such conditions are only needed in establishing the detailed structure of these regions, which is directly related to the fact that the finite front-thickness effects determining this structure are described by equations of higher differential order.
This means that the slope of the function $f(x)$ describing the outer solution is to be considered large everywhere on the semi-open intervals $x \in (0,1]$ and $x \in [-1,0).$ The points $x = \pm 1$ are included here by continuity, but not the point $x=0$ which represents the true singularity of the flow, where $f'$ is undefined.

\section{Solution of Eq.~(\ref{generalc1st}) in the high-velocity limit}\label{solution}

\subsection{Large-slope expansion of the $\EuScript{H}$-operator}\label{largesl}

The nature of nonlocality of steady flames is encoded entirely in the structure of the operator $\hat{\EuScript{H}},$ and of major importance is the fact that this operator greatly simplifies in the high-velocity limit. For large $U,$ the front
slope is also large, $|f'|\sim U,$ so the argument of cotangent in
Eq.~(\ref{hcurvedf}) has a large imaginary part for almost all
values of the integration variable. Therefore, one can write
\begin{eqnarray}\label{approxcot}
\cot\left\{\frac{\pi}{2}\left(\eta - x + i[f(\eta) -
f(x)]\right)\right\} \approx - i\chi(|\eta| - |x|)\,,
\end{eqnarray}
\noindent where $\chi(x)$ is the sign function,
$$\chi(x) = \left\{
\begin{array}{cc}
+1,& x>0\,,\\
-1,&  x<0\,.
\end{array}
\right.
$$ This approximation is valid for all $\eta$ except two
small regions near $\eta = \pm|x|\,.$ More precisely, taking
into account that, for real $a_{1,2},$ $$\cot(a_1+ia_2) =
-i\,\frac{e^{(a_2 - ia_1)} + e^{-(a_2 - ia_1)}}{e^{(a_2 - ia_1)} -
e^{-(a_2 - ia_1)}} = -i\chi(a_2) + O\left(e^{-2|a_2|}\right)\,,$$ we
see that Eq.~(\ref{approxcot}) holds true, with an exponential
accuracy, everywhere except $$\eta: |\eta| \in (|x| -
\delta, |x| + \delta),$$ where $\delta = O(1/U).$

To develop an asymptotic large-slope expansion of $\hat{\EuScript{H}},$ let us choose a real $\varepsilon > 0$
such that
\begin{eqnarray}\label{epsiloncond}
\varepsilon \ll 1\,, \quad U\varepsilon \gg 1\,.
\end{eqnarray}
\noindent Then the integral in Eq.~(\ref{hcurvedf}) can be
rewritten, for $x>0,$ as
\begin{eqnarray}\label{intred}
\fint\limits_{-1}^{+1}
&&d\eta~a(\eta)\cot\left\{\frac{\pi}{2}\left(\eta - x + i[f(\eta) -
f(x)]\right)\right\} \nonumber\\ = &&-
i\left[\int\limits_{-1}^{-x-\varepsilon} +
\int\limits_{-x+\varepsilon}^{0} + \int\limits_{0}^{x-\varepsilon} +
\int\limits_{x+\varepsilon}^{+1}\right]
d\eta~a(\eta)\chi(|\eta| - x) \nonumber\\&& +
\left[\int\limits_{-x-\varepsilon}^{-x+\varepsilon} +
\fint\limits_{x-\varepsilon}^{x+\varepsilon}\right]
d\eta~a(\eta)\cot\left\{\frac{\pi}{2}\left(\eta - x + i[f(\eta) - f(x)]\right)\right\}\,.
\end{eqnarray}
\noindent Notice that in the last term on the right hand side of Eq.~(\ref{intred}), only one of the two integrals is defined in the principal value sense. As
such, it is proportional to $a'(x).$ It is not difficult to see that contributions of this kind give rise to terms of the order $1/U^2.$ This is because expanding the function $a(\eta)$ around $x$ brings in an extra small factor $(\eta - x).$ Below, we will need $\hat{\EuScript{H}}$ expanded up to $O(1)$-terms, so the principal-sense integral can be neglected. The other integral can be evaluated as follows, using continuity of $a(\eta)$
\begin{eqnarray}&&
\int\limits_{-x-\varepsilon}^{-x+\varepsilon}d\eta~a(\eta)
\cot\left\{\frac{\pi}{2}\left(\eta - x + i[f(\eta) - f(x)]\right)\right\}
= - i a(-x) \int\limits_{-\varepsilon}^{+\varepsilon}dy\coth \left\{\frac{\pi f'(x)}{2}y + \pi ix\right\} \nonumber\\&& =
- i a(-x)\left.\frac{2}{\pi f'(x)} \ln{\rm sh} y\right|_{-\pi f'(x)\varepsilon/2 + \pi i x}^{+\pi f'(x)\varepsilon/2 + \pi i x}\,.\nonumber
\end{eqnarray}
\noindent  In view of (\ref{epsiloncond}), $\ln{\rm sh} y$ can be replaced, with the exponential accuracy, by $y$ and $-y$ at the upper and lower integration limits, respectively. Taking into account also that for $x\to 0^+,$ ${\rm arg}({\rm sh} y)$ gains $-\pi,$ we find
\begin{eqnarray}&&
\left.\phantom{\frac{2}{\pi f'(x)}} \ln{\rm sh} y\right|_{-\pi f'(x)\varepsilon/2 + \pi i
x}^{+\pi f'(x)\varepsilon/2 + \pi i x} = \pi
i(2x - 1)\,.\nonumber
\end{eqnarray}
\noindent On the other hand, replacing the cotangent in the last term in Eq.~(\ref{intred}) by the sign function gives zero within the same accuracy
\begin{eqnarray}&&
\int\limits_{-x-\varepsilon}^{-x+\varepsilon}d\eta~a(\eta)
\chi(|\eta| - x) = a(-x)
\int\limits_{-\varepsilon}^{+\varepsilon}d\eta~\chi(\eta)
= 0\,.\nonumber
\end{eqnarray}
\noindent Using these formulas in Eq.~(\ref{intred}), and then
putting it in Eq.~(\ref{hcurvedf}) gives finally
\begin{eqnarray}\label{hcurvedf2}
\left(\hat{\EuScript{H}}a\right)(x) = (f'(x) -
i)~\int\limits_{0}^{+1} d\eta~\frac{a(\eta) +
a(-\eta)}{2}\chi(\eta - |x|)+ia(-x)(2|x| - 1) +
O\left(\frac{1}{U}\right)\,.\nonumber\\
\end{eqnarray}
\noindent This result is written in the form applicable to negative as well as positive $x,$ which can be verified by noting that $i\hat{\EuScript{H}}$ is invariant under the combined operation of coordinate inversion $(x\to -x)$ and complex conjugation, as is seen from Eq.~(\ref{hcurvedf}). In particular, the asymptotic action of $\hat{\EuScript{H}}$ on the derivative of a function $a(x)$ continuous at the origin and satisfying $a(-1) = a(+1),$ is
\begin{eqnarray}\label{hcurvedf3}
\left(\hat{\EuScript{H}}a'\right)(x) = (f'(x) -
i)\left\{a(-|x|) - a(|x|)\right\} + ia'(-x)(2|x| - 1) +
O\left(\frac{1}{U}\right),
\end{eqnarray}
\noindent where the prime now denotes derivative of the function
with respect to its argument, $a'(y) = da(y)/dy.$ It turns out, however, that the formula (\ref{hcurvedf3}) remains valid even if the conditions $a(0^-) = a(0^+),$ $a(-1) = a(+1)$ are not met. In particular, $a(x)$ can be discontinuous at $x=0,$ so that its derivative is singular at the origin. This important fact is proved in the appendix.

Formula (\ref{hcurvedf3}) was derived for $x$'s where the front slope is large, {\it i.e.,} for all $x$ except small regions near the rod and the channel walls, where front curvature is large. Neglecting these regions as we did before, we can say that Eq.~(\ref{hcurvedf3}) is valid on the semi-open intervals $x\in (0,+1]$ and $x\in [-1,0)$ (see Sec.~\ref{evolutionequation}).

The following comments concerning the structure of Eqs.~(\ref{hcurvedf2}), (\ref{hcurvedf3}) will be useful in subsequent applications. First, it is
seen that the result of the action of $\hat{\EuScript{H}}$ depends
essentially on parity properties of the function $a(x),$ namely,
$\hat{\EuScript{H}}a = O(U),$ if $a(x)$ is even, and
$\hat{\EuScript{H}}a = O(1),$ if it is odd. Second, it should be noted
that although the identity $\hat{\EuScript{H}}^2 = - 1$ is valid whatever the shape of the flame-front, in particular, in the large-$U$ limit, it cannot be
verified using the expression on the right of Eq.~(\ref{hcurvedf2}),
already because the composition of its leading term with the
undetermined remainder $O(U)\circ O(1/U) = O(1).$

\subsection{Reduction to the system of ordinary differential equations}\label{reduction1}

\subsubsection{Equation for the transversal velocity component}

We now go over to proving the results announced in Sec.~\ref{strategy}. First off all, let us determine the orders of various terms in Eq.~(\ref{generalc1str}) within the high-velocity expansion. As we know, $u_- = O(U),$ $f'=O(U),$ so Eq.~(\ref{evolutiongen1}) tells us that $w_- = O(1).$ Using these estimates in the expressions (\ref{jumps}), (\ref{vorticity}) shows that
$$[u] = O(1/U)\,, \!\!\quad [w] = O(1)\,, \!\!\quad u_+ = O(U)\,, \!\!\quad w_+ = O(1)\,, \!\!\quad v_+ = O(U)\,, \!\!\quad N\sigma_+ = O(U^2)\,.$$ Therefore, one has for the braces in Eq.~(\ref{generalc1str})
$${\rm Re}\left\{[\omega]' -
\frac{Nv^n_+\sigma_+\omega_+}{v^2_+} \right\} = O(U)\,, \quad {\rm Im}\left\{[\omega]' -
\frac{Nv^n_+\sigma_+\omega_+}{v^2_+} \right\} = O(1)\,.$$ Since ${\rm Re}\{...\}$ is an odd function of $x,$ while ${\rm Im}\{...\}$ is even, formula (\ref{hcurvedf2}) shows that the leading term in the real part of $i\hat{\EuScript{H}}\{...\}$ is $O(U)\,.$ Thus, the real part of the left hand side of Eq.~(\ref{generalc1str}) is $O(U).$ At the same time, as we saw in Sec.~(\ref{boundarycond}), the real part of the dipole velocity field is $O(1)$ in the matching region, and negligible far apart from the origin. Therefore, to the leading order of the high-velocity expansion, the real contribution to the right hand side of Eq.~(\ref{generalc1str}) can be omitted.
However, things are quite different for the imaginary contribution. In this case,
the above estimates and formula (\ref{hcurvedf2}) show that both sides of Eq.~(\ref{generalc1str}) are $O(1).$ Moreover, expansion of $\hat{\EuScript{H}}$ only up to $O(1)$-terms is actually insufficient for the purpose of extracting the imaginary part. Indeed, the undetermined remainder in Eq.~(\ref{hcurvedf2}) is $O(1/U),$ and may contain real as well as imaginary parts. Since the argument of the $\EuScript{H}$-operator is $O(U),$ this remainder gives rise to terms of the order $O(1/U)\cdot O(U) = O(1).$ These two complications can be overcome by resorting to Eq.~(\ref{chup}) which expresses potentiality of the upstream flow, and can be considered as a consistency condition for Eq.~(\ref{generalc1str}). Repeating literally the above arguments,\footnote{For a flame stabilized in the wake,  Eq.~(\ref{chup}) is replaced by $\left(1 - i \hat{\EuScript{H}}\right)\left(\omega_-\right)' =
2\left(\omega^d_-\right)'$ \citep{jerk3}. This makes no difference for the present analysis, as the dipole field is negligible on the same grounds as in Eq.~(\ref{generalc1str}). \label{footn1}} one sees that the real part of Eq.~(\ref{chup}) can be consistently extracted with the help of the expansion obtained in the previous section. Namely, using the formula (\ref{hcurvedf2}) yields
$$u_-'(x) - f'(x)\{w_-(|x|) - w_-(-|x|)\} + u_-'(-x)(2|x| - 1) = 0\,,$$ or since $u_-'(-x) = - u_-'(x),$ $w_-(-x) = - w(x),$
\begin{eqnarray}\label{chup1}
w_-(|x|) = (1 - |x|)\frac{u_-'(x)}{f'(x)}\,.
\end{eqnarray}
\noindent By the construction, this equation is valid for $x\in [-1,0)\cup (0,1].$ In particular, we see that the boundary condition (\ref{bcpsirw1}) is satisfied automatically.

Thus, we proved that in order to find the large-scale solutions of Eq.~(\ref{generalc1str}), it is sufficient to consider Eq.~(\ref{generalc1st}).

\subsubsection{Equation for the longitudinal velocity component}\label{eqlong}

\begin{figure}
\centering
\includegraphics[width=.8\textwidth]{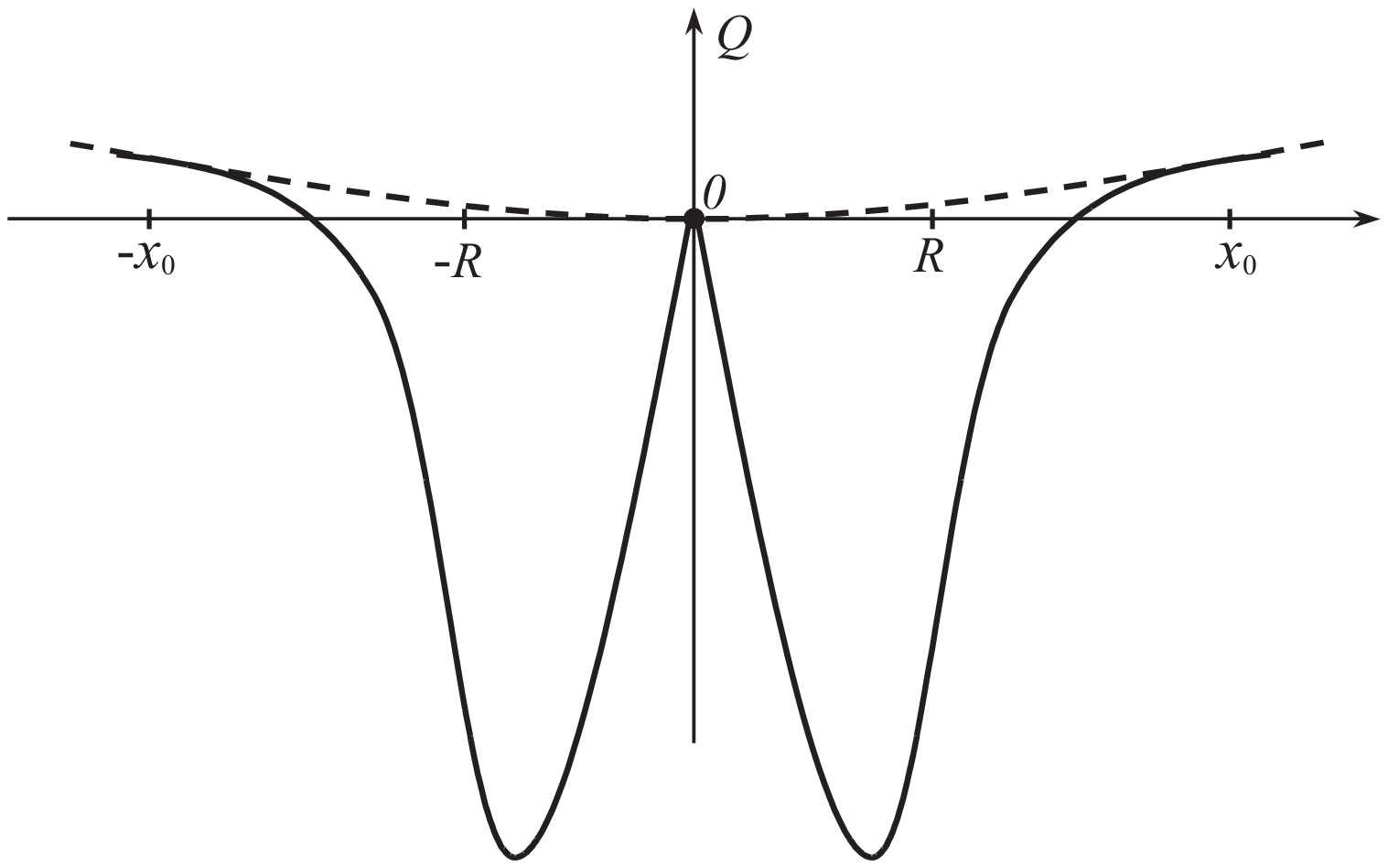}
\caption{}\label{fig3}
\end{figure}

Turning back to extracting the real part of Eq.~(\ref{generalc1st}), we have to consider the question concerning the contribution of the small region near the rod. As we saw in the preceding section, the dipole field on the right of Eq.~(\ref{generalc1str}) can be neglected for $x\ne 0.$ However, this does not settle the question, because integration on the left hand side extends over all $x$ including zero. The first term in the braces is a derivative, so that we can use Eq.~(\ref{hcurvedf3}) to find how it is transformed by $\hat{\EuScript{H}}.$ Note that since this term is the derivative of a function discontinuous at $x=0,$ it contains contribution proportional to the Dirac $\delta$-function. Indeed, one has from Eq.~(\ref{jumps})
$$[\omega] = \frac{\theta - 1}{N}
- if'(x)\frac{\theta - 1}{N} \approx - i(\theta - 1)\chi(x)\,,$$ so that
$$[\omega]' = - 2i(\theta - 1)\delta(x)\,.$$ However, it is proved in the Appendix that the formula (\ref{hcurvedf3}) is still applicable in this case. As to the second term in the braces, it also contains a derivative factor, {\it viz.,} $(u^2_- + w^2_-)'$ coming from $\sigma_+,$ but this time this is a derivative of an even function. However, it would be premature to conclude that this term does not contain a $\delta$-contribution. The point is that $(u^2_- + w^2_-)'$ is multiplied by $\omega_+$ whose imaginary part is an odd function. Let us trace the development of the quantity $Q = Nw_+\sigma_+/v^2_+$ near the rod (see Fig.~\ref{fig3}). At the matching point $x = - x_0$ on the left of the rod, $v^2_+$ is large but changes slowly, while $w_+ = -w_0 + (\theta-1) = O(1).$ Also, if the bulk transversal velocity of fresh gas is not too large, and $|v_+|$ increases away from the anchor (pressure normally drops down along the stream), then $w_+>0,$ $(v^2_+)'<0,$ and so
\begin{eqnarray}\label{q}
0 < Q = - w_+\frac{(\theta - 1)}{2\theta}\left(\ln v^2_+\right)' = O(1)\,.
\end{eqnarray}
\noindent
$|\sigma_+|$ increases near the rod as the result of gas slowdown caused by the rod,  and for $|x|\lesssim R,$ $Q$ becomes $O(U/R).$ Here, the front curvature is large, and the zero-front-thickness expression (\ref{q}) can be used only for a rough estimate. It shows that $Q$ is negative in this region, because $w_+<0,$ and $\sigma_+>0.$ $Q$ rapidly turns into zero at $x=0,$ because both $w_+$ and $\sigma_+$ vanish at the origin. For positive $x \lesssim R,$ $Q$ is again a negative $O(U/R)$ quantity, since $w_+>0,$ $\sigma_+<0.$ At larger $x$'s its modulus decreases, and $Q$ becomes $O(1)$ at the matching point $x = x_0$ on the right of the rod. From the large-scale point of view, this behavior means that $Q$ contains a term $q\delta(x),$ with a negative coefficient $q.$ The exact value of $q$ can be found, of course, only if the inner solution is known. We arrive at the conclusion that the expression in the braces in Eq.~(\ref{generalc1st}) can be written as
$$- i\bar{q}\delta(x) + (\theta - 1)\frac{(u^2_- + w^2_-)'\omega_+}{2v^2_+}\,,$$
where $\bar{q} = q + 2(\theta - 1),$ and it is understood that the $\delta$-contribution is excluded from the second term. In other words, this term is calculated using the functions $u_-,w_+$ etc. that describe the outer solution.  Taking into account that $v^2_+ = v^2_- + \theta^2 - 1,$ using Eqs.~(\ref{hcurvedf2}), (\ref{hcurvedf3}), and extracting the real part of Eq.~(\ref{generalc1st}) gives, to the leading order,
$$u'_-(x) (1 + \alpha |x|) - \frac{\bar{q}}{2}f'(x)
- \frac{\alpha}{2}f'(x)\int\limits_{0}^{1}d\eta \frac{u'_-(\eta)}{u_-(\eta)}[w_-(\eta) - \alpha]\chi(\eta - |x|) = 0\,, \quad \alpha \equiv \theta - 1\,.$$
This equation involves the unknown parameter $\bar{q}.$ To get rid of it, we divide  the equation by $f',$ and then differentiate it with respect to $x.$ The result is the following ordinary differential equation
\begin{eqnarray}\label{main1}
\frac{d}{dx}\left[\frac{u'_-(x)}{f'(x)} (1 + \alpha |x|)\right] + \alpha\frac{u'_-(x)}{u_-(x)}[w_-(x) - \alpha] = 0\,.
\end{eqnarray}
\noindent
Together with Eqs.~(\ref{evolutiongen1}), (\ref{chup1}) it constitutes the system of three ordinary differential equations for the three functions $u_-(x),w_-(x),f(x).$ Evidently, it requires three initial conditions which are Eqs.~(\ref{bcpsir1}) -- (\ref{bcpsirf}).

\subsection{Reduction to a single differential equation. Numerical solutions}\label{reduction}

Introducing an auxiliary function $$\varphi = \frac{u'_-}{f'} = \frac{d u_-}{d f}\,,$$ the system (\ref{evolutiongen1}), (\ref{chup1}), (\ref{main1}) can be rewritten as an ordinary differential equation for $\varphi(x)$:
\begin{eqnarray}\label{main2}
\frac{d}{dx}\left[\varphi (1 + \alpha x)\right] + \alpha\varphi\frac{(1 - x)\varphi - \alpha}{(1 - x)\varphi + 1} = 0\,, \quad x>0\,.
\end{eqnarray}
\noindent The initial condition for $\varphi$ follows from Eqs.~(\ref{bcpsir1}), (\ref{chup1}): $\varphi(0^+) = w_0.$
We note also that the functions $u_-(x),$ $f(x)$ are related by a simple algebraic equation. One has from Eqs.~(\ref{evolutiongen1}), (\ref{chup1})
\begin{eqnarray}\label{uf}
u_- = (1-x)u'_- + f'\,,
\end{eqnarray}
\noindent
or $$[(1-x)u_-]' + f' = 0\,.$$ Integrating this equation, and using the initial conditions (\ref{bcpsiru1}), (\ref{bcpsirf}) gives
\begin{eqnarray}\label{main3}
f = U - (1 - x)u_-\,.
\end{eqnarray}
\noindent Finally, combining Eqs.~(\ref{uf}), (\ref{main3}) one can express  $\varphi$ in terms of $f$
$$\varphi(x) = \frac{U - f(x)}{(1 - x)^2 f'(x)} - \frac{1}{1 - x}\,.$$ Substitution of this expression into Eq.~(\ref{main2}) leads to a second-order differential equation for the front position. The corresponding initial conditions follow from Eqs.~(\ref{bcpsir1}) -- (\ref{bcpsirf}), and (\ref{evolutiongen1}):
\begin{eqnarray}\label{initc}
f(0) = 0\,, \quad f'(0^+) = \frac{U}{w_0 + 1}\,.
\end{eqnarray}
\noindent

\begin{figure}
\includegraphics[width=.4\textwidth]{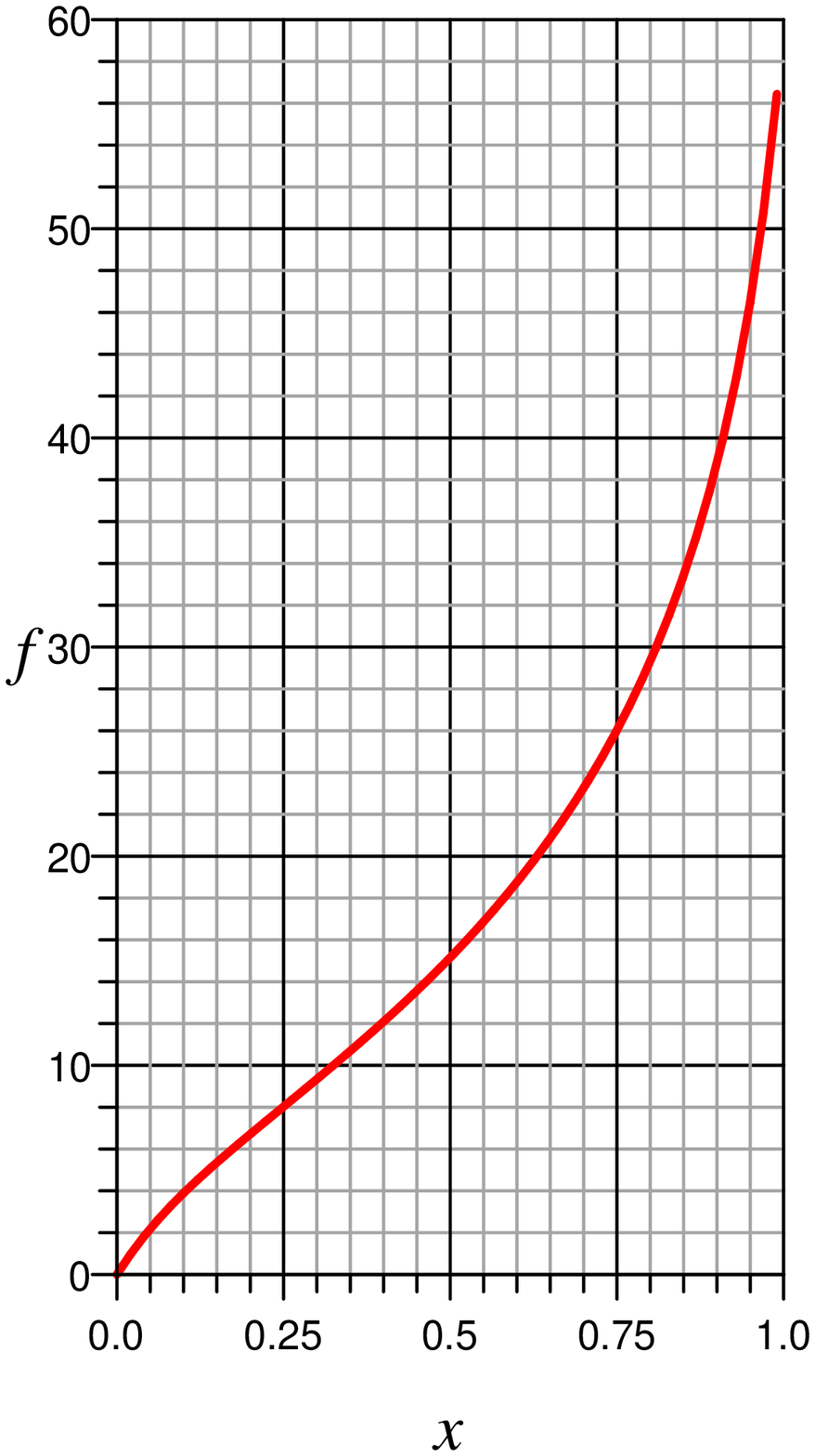}\\
\includegraphics[width=.33\textwidth]{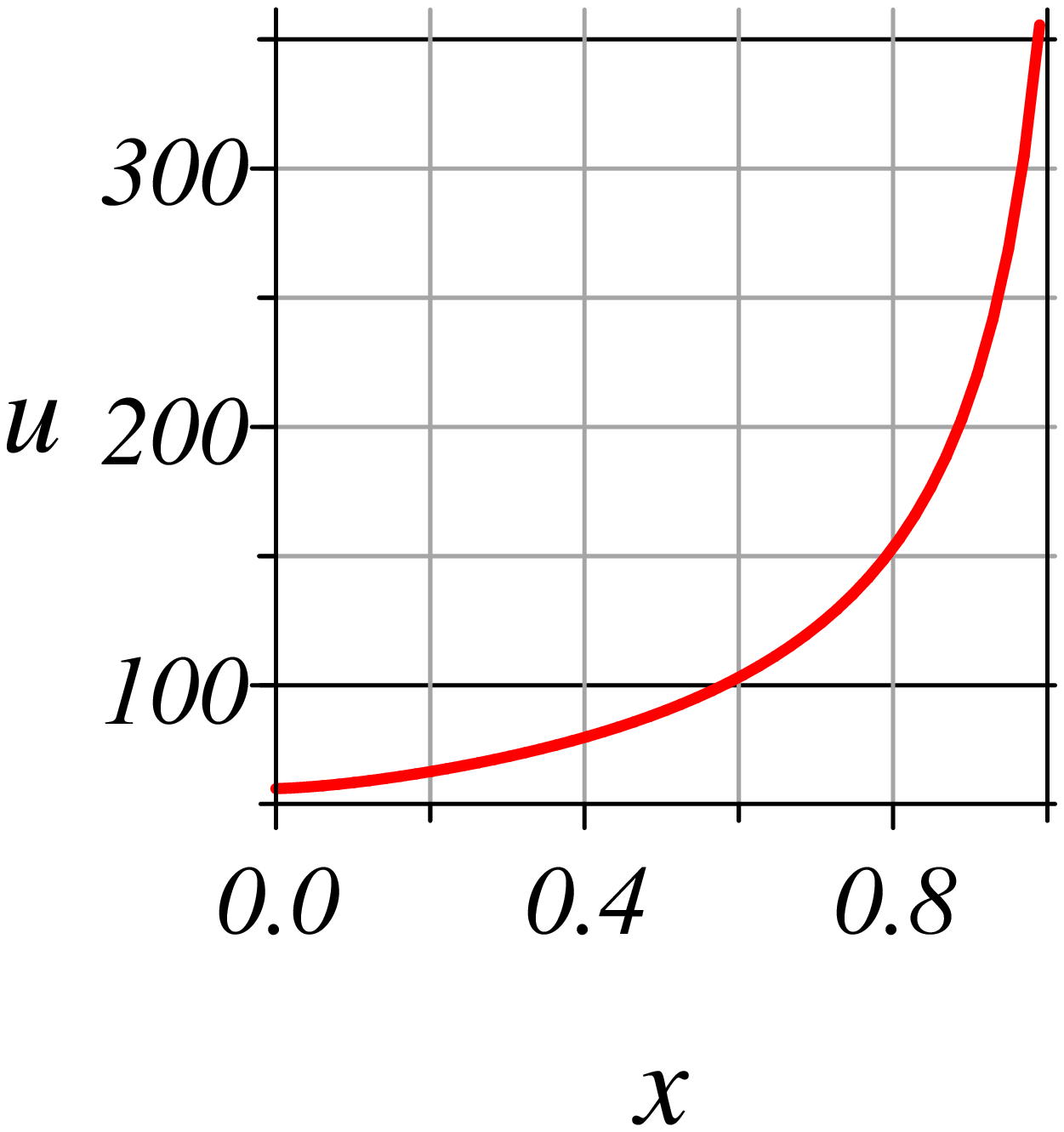}\hspace{2cm}
\includegraphics[width=.3\textwidth]{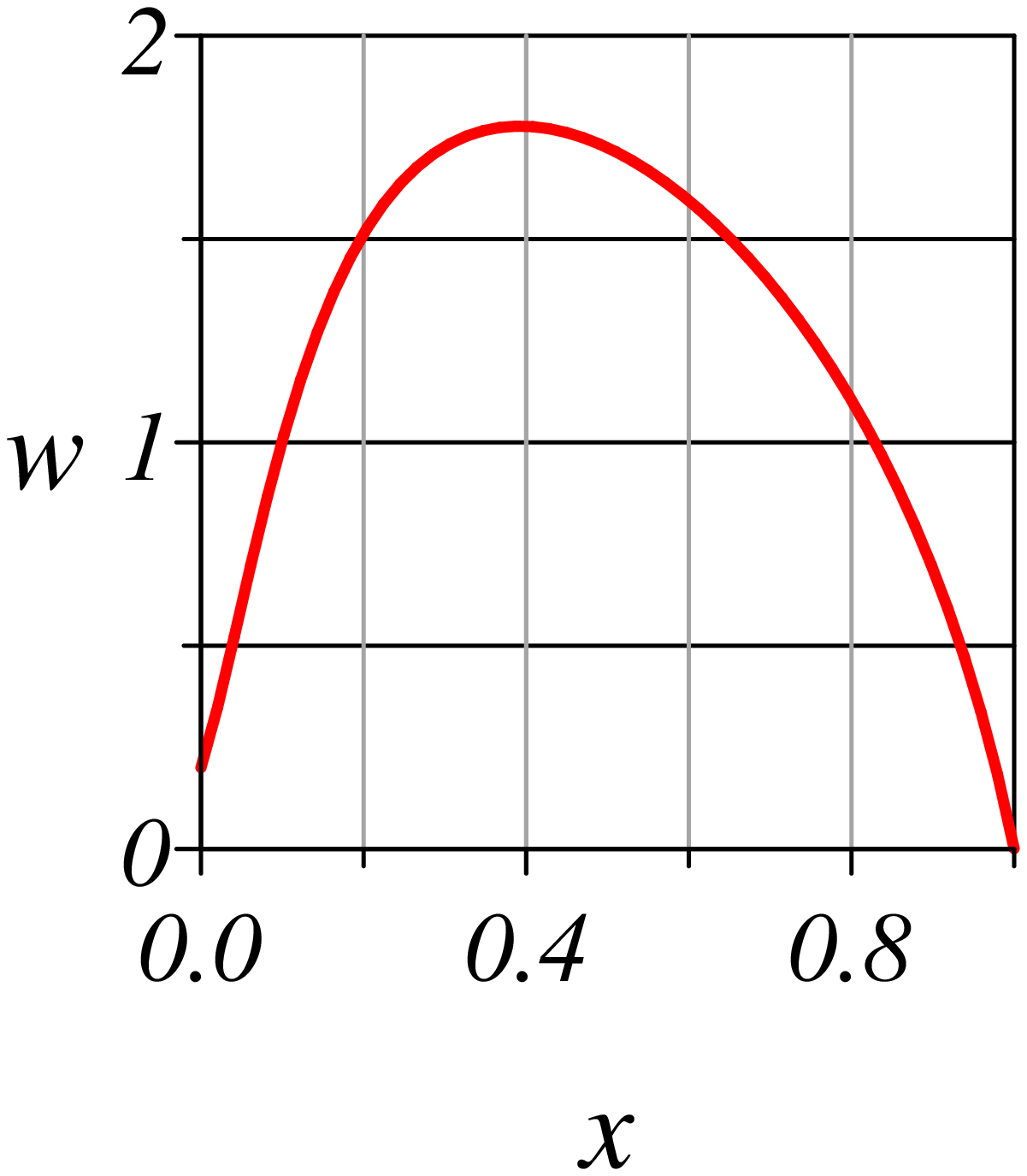}
\caption{}\label{fig4}
\end{figure}

\begin{figure}
\centering
\includegraphics[width=.4\textwidth]{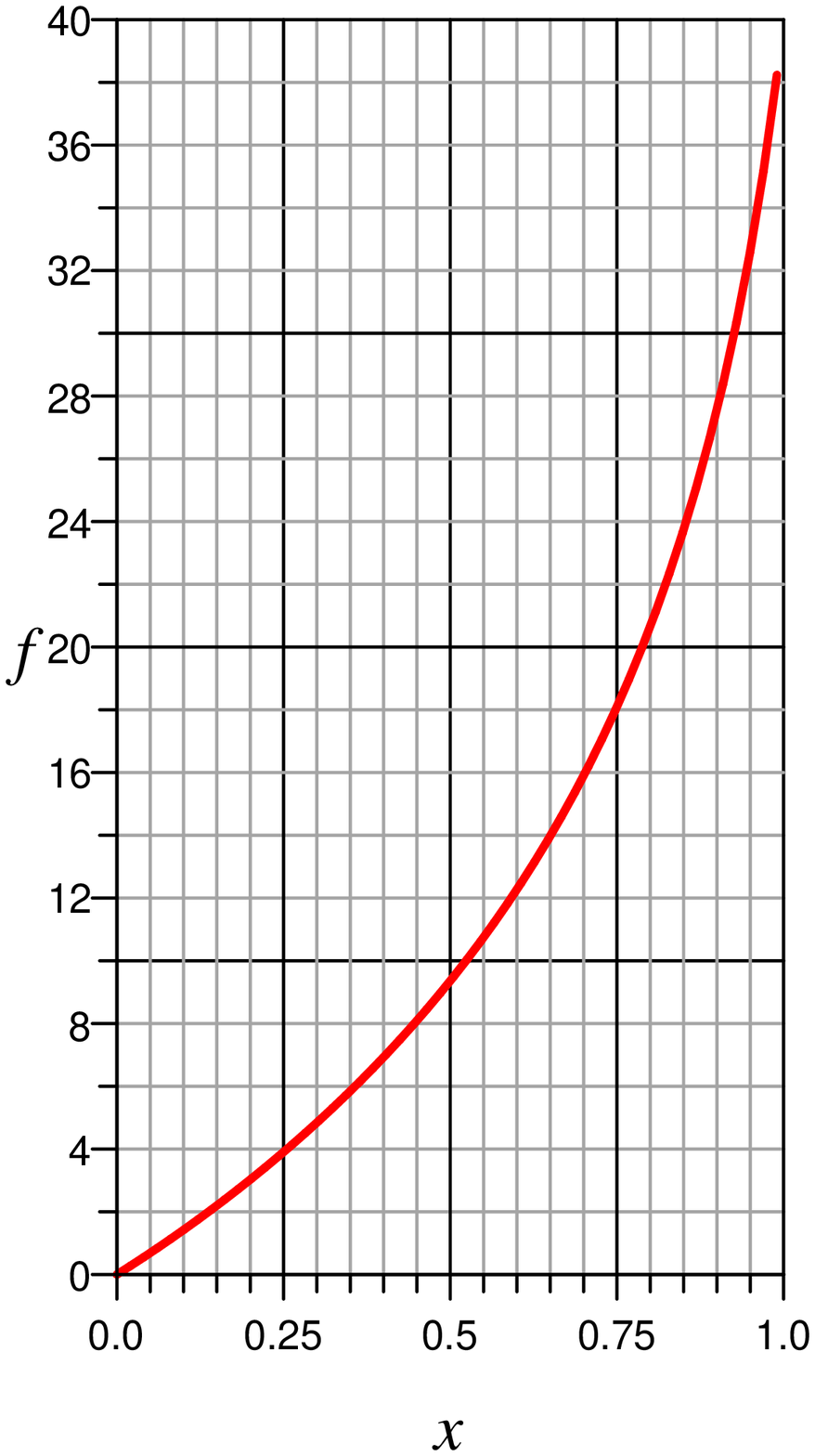}\\\vspace{1cm}
\includegraphics[width=.31\textwidth]{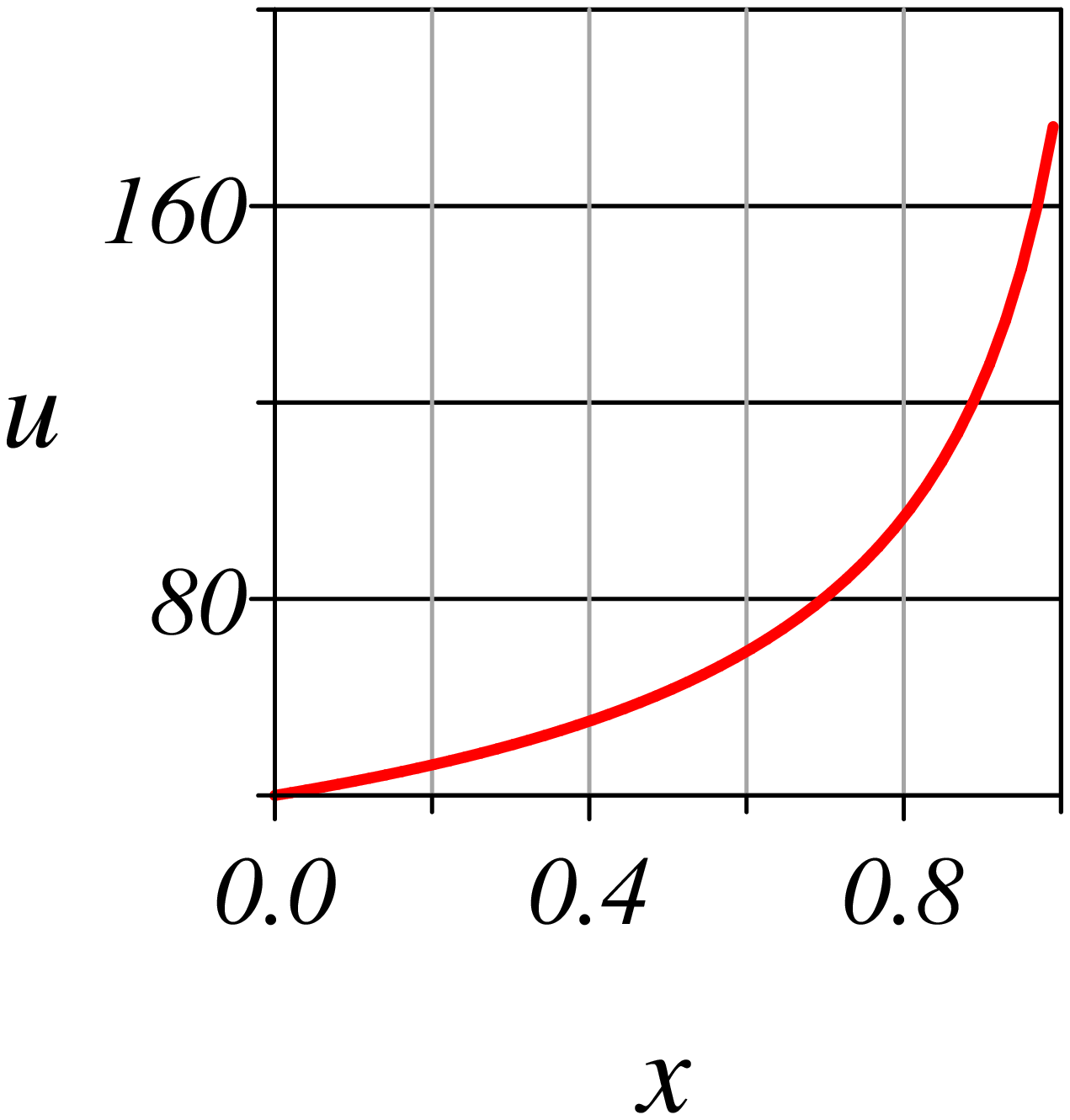}\hspace{3cm}
\includegraphics[width=.3\textwidth]{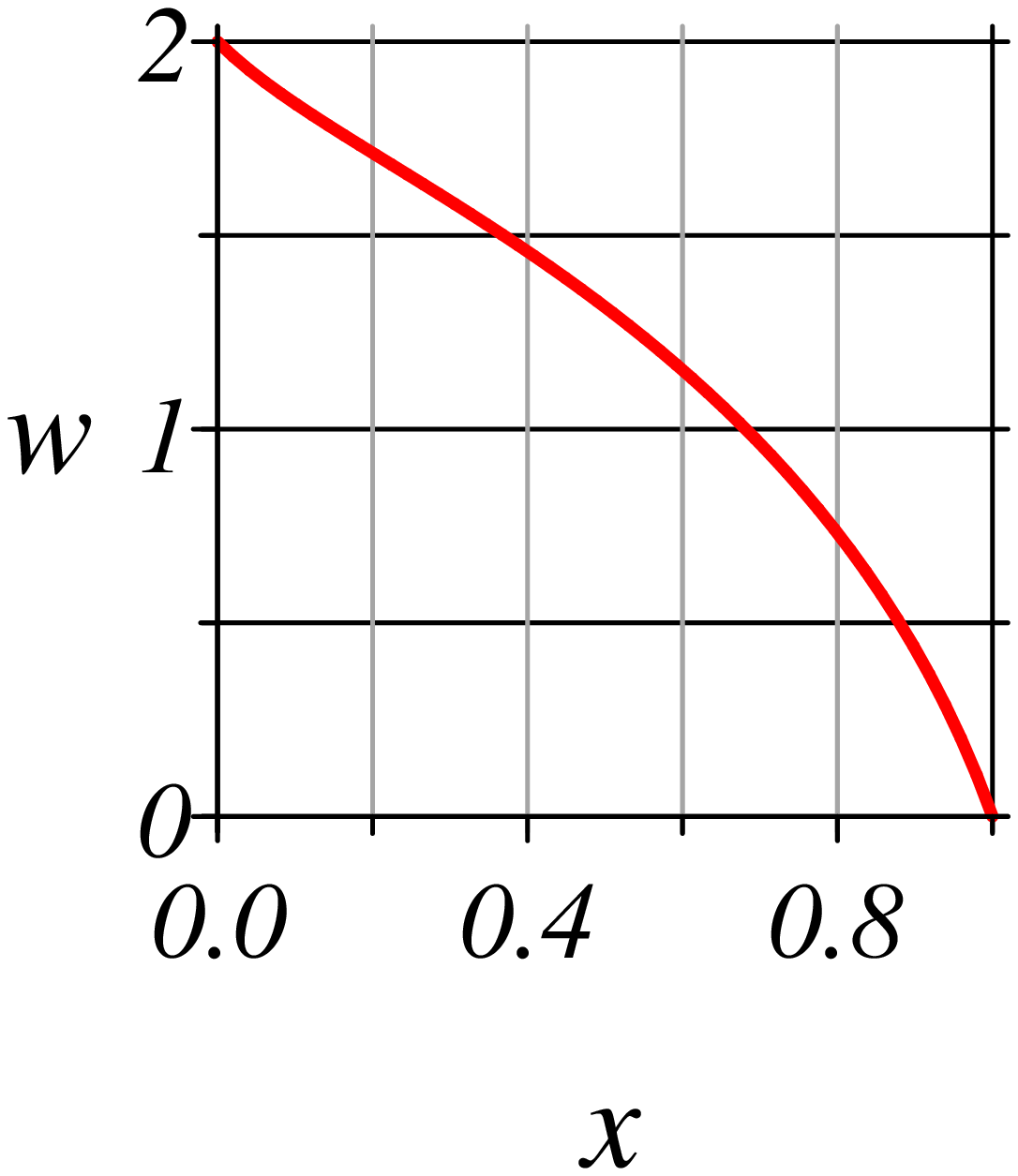}
\caption{}\label{fig5}
\end{figure}

\begin{figure}
\centering
\includegraphics[width=.4\textwidth]{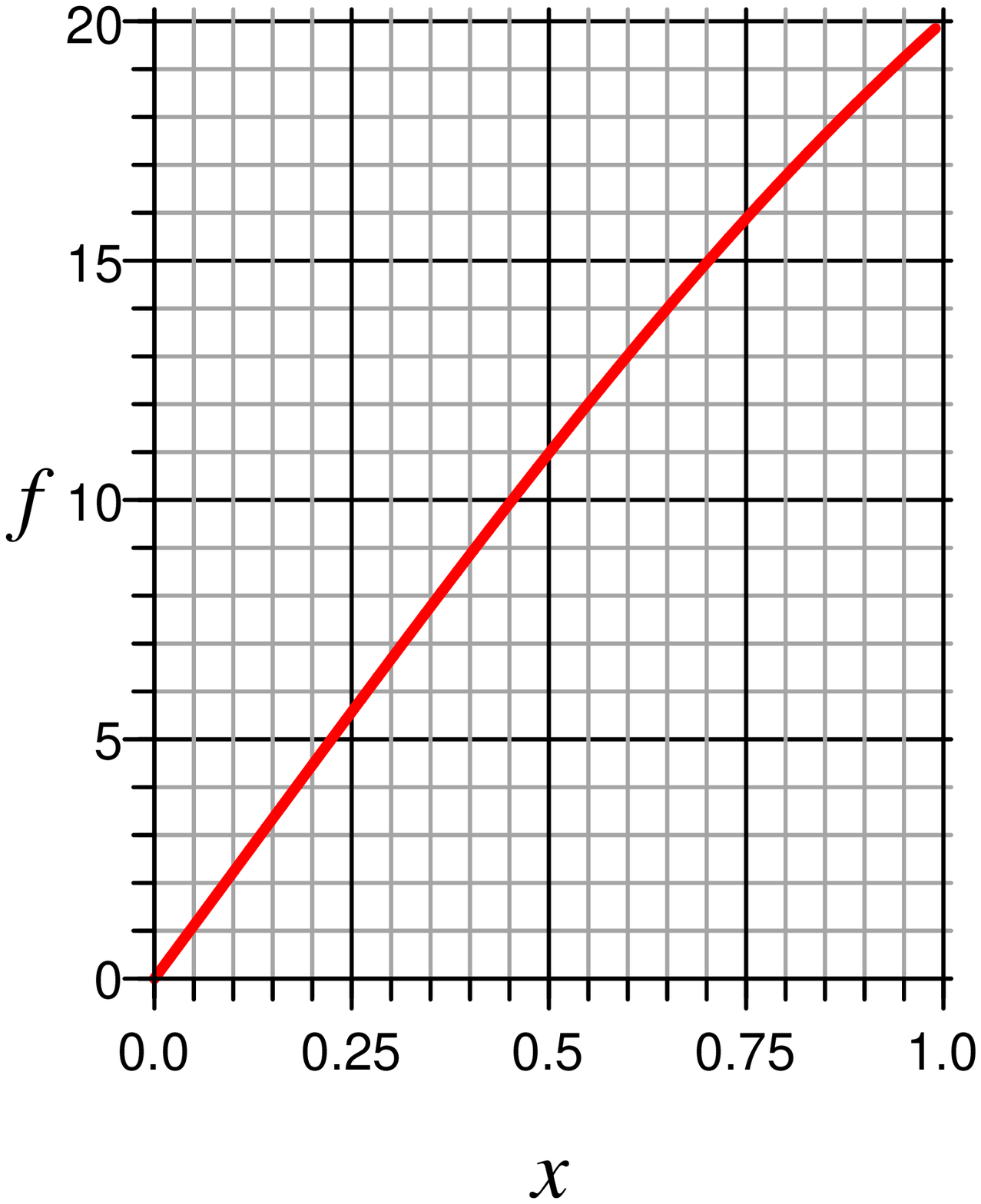}\\\vspace{1.5cm}
\includegraphics[width=.3\textwidth]{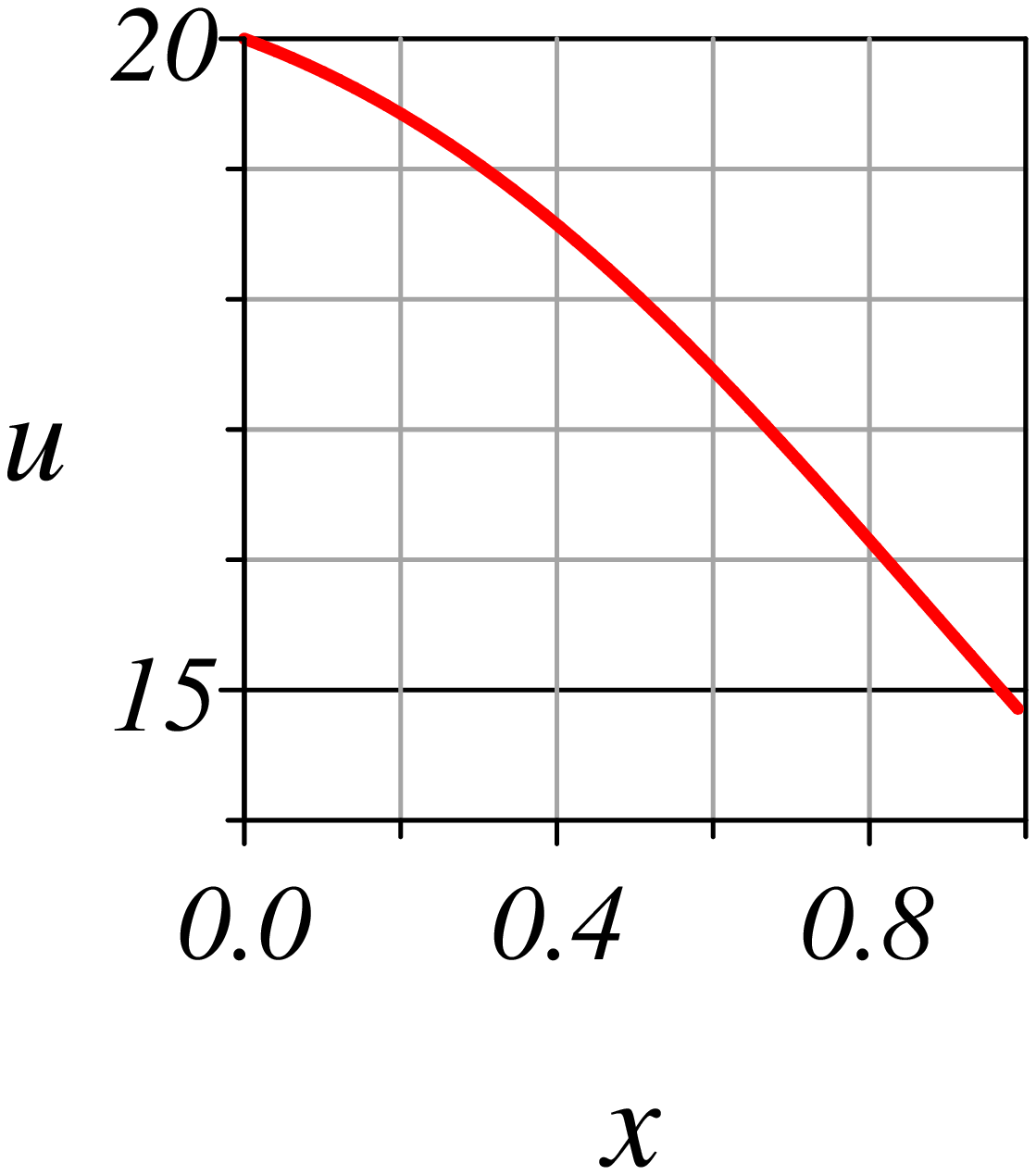}\hspace{2cm}\vspace{-1cm}
\includegraphics[width=.33\textwidth]{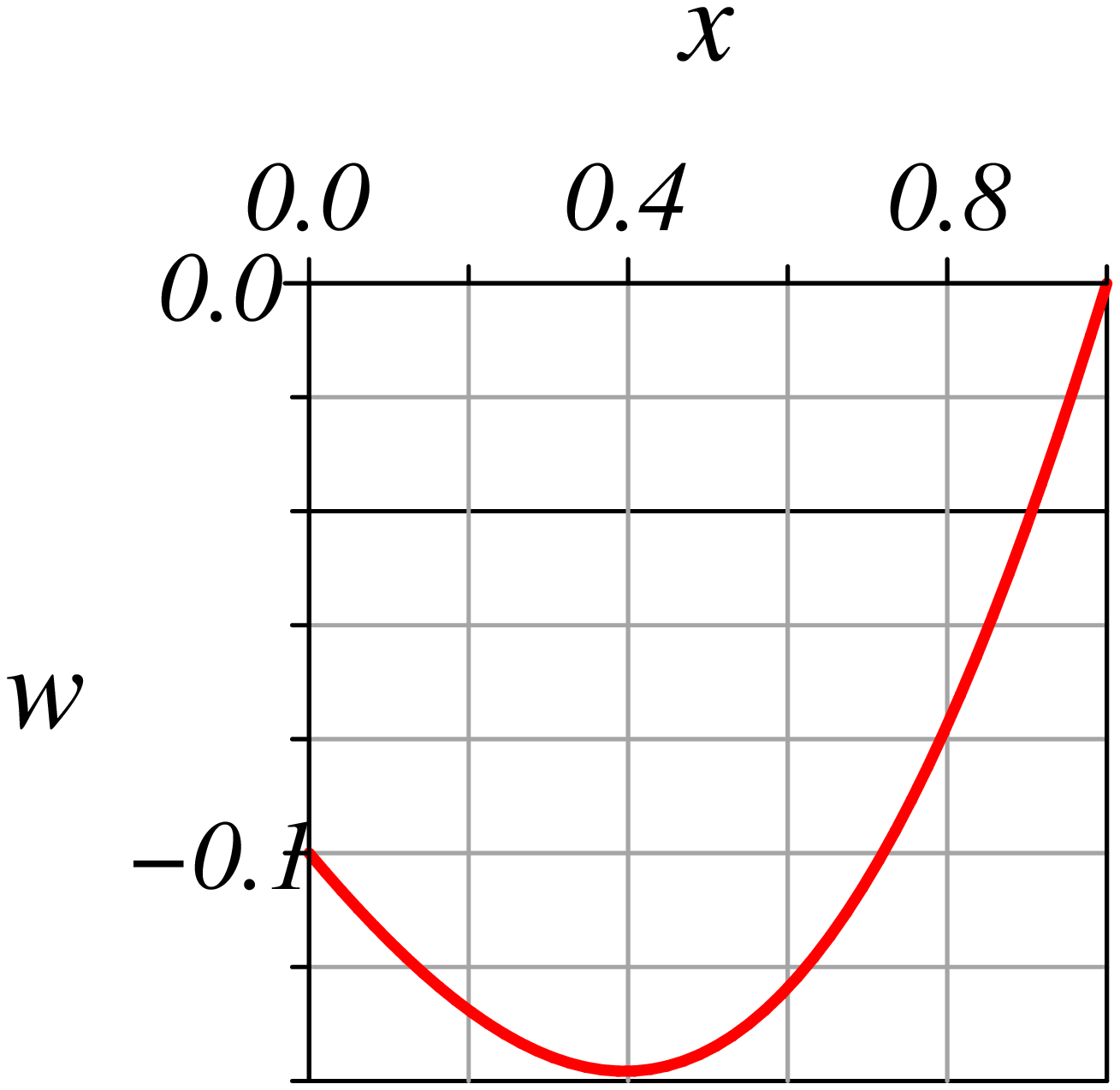}
\caption{}\label{fig6}
\end{figure}

Numerical solutions of these equations for various values of $U,w_0$ and $\theta$ are plotted in Figs.~\ref{fig4} -- \ref{fig6}. They have the following general features. First of all, the function $f(x)$ is monotonic in all cases (in each of the two channel halves), as was assumed throughout our consideration. Second, flames in which the fresh-gas flow diverges near the rod ($w_0 >0$) are convex towards the incoming flow, while those with convergent fresh-gas flow ($w_0<0$) are concave (recall that we deal here with the large-scale solutions, characterized by distances $r\gtrsim R_0$ from the rod; for $|x|<R,$ of course, $w_0$ is positive in any case). Numerical analysis shows that in the latter case, solutions with the given negative $w_0$ exist only for sufficiently small values of the gas expansion coefficient. For example, solutions with the fairly small value $w_0 = - 0.1,$ one of which is shown in Fig.~\ref{fig6}, disappear at $\theta \approx 3.5,$ and this threshold value is independent of $U.$ Finally, solutions with $w_0 >0$ are characterized by monotonic increase of $u_-$ as one moves from the rod to the wall. The overall velocity rise is substantial -- typically two to four times. This is what normally observed in experiments. Solutions with $w_0<0$ are anomalous in this respect, as $u_-$ slightly decreases away from the rod. They are most likely unstable.

At last, solutions with $w_0=0$ are trivial. Indeed, in this case $\varphi(0^+) = 0,$ and Eq.~(\ref{main2}) tells us that also $\varphi'(0^+)=0.$ Repeated differentiation of this equation then shows that all higher derivatives of $\varphi$ also vanish, {\it i.e.,} $\varphi(x)\equiv 0.$ Hence, $u_- = {\rm const} = U,$ and Eq.~(\ref{main3}) gives $f(x) = Ux,$ $x>0.$ In the light of the discussion given in Sec.~\ref{boundarycond}, it is natural that the case $w_0=0$ reproduces the result of Sec.~\ref{chernyf}.

\section{Conclusions}\label{conclusions}

The results obtained in this paper provide consistent description of steady anchored flames in high-velocity gas streams in channels. Given the values of the incoming flow velocity and its transversal component near the anchor, the formulas derived in Sec.~\ref{reduction} allow simple determination of the flame front shape and on-shell gas-velocity. A practically more convenient may be ``geometrical'' parametrization using the ordinates of the front end-points and its slope at the origin, which is related to the initial one by Eqs.~(\ref{lu}), (\ref{initc}).

The remarkable fact revealed by the above investigation is that the flame structure in a high-velocity gas flow obeys an ordinary differential equation. In other words, this structure turns out to be local in the usual sense: behavior of the flame front slope and gas velocity in an infinitesimal vicinity of a given point is determined by their values at this point. The sole role of the anchor is to provide an initial condition. This result answers the question as to the nature of nonlocality of the anchor influence on the flame structure: Although detailed structure of the flame holder is immaterial for the properties of the large-scale flow, the flow distortion it causes ultimately determines the whole flame configuration. We saw in Sec.~\ref{boundarycond} that mathematically, the presence of the anchor with vanishingly small dimensions signifies existence of a singularity in the bulk flow solution. The failure to recognize this fact is what makes it impossible to consistently describe nontrivial flame configurations within the Zel'dovich-Scurlock-Tsien approach. Actually, this defect is inherent to this approach as it neglects the transversal gas velocity, while the role of this component is crucial in describing the anchor impact.

Finally, the role of vorticity in the formation of curved flames is to be emphasized. It is described by the second term in the braces in Eq.~(\ref{generalc1st}), while the first term (the complex velocity jump) corresponds to a purely potential contribution. As we have seen in Sec.~\ref{eqlong}, the latter falls off from the equations describing the large-scale flame structure. One can say that formation of the steady flame pattern in a high-velocity stream is governed by the vorticity generated in the curved flame front. Therefore, it cannot be described within potential-flow models such as suggested in \citep{frankel1990}.

\acknowledgments{I am grateful to Guy Joulin and Hazem El-Rabii for discussions of various issues considered in the paper. Although this work was not discussed directly with my colleagues, our numerous conversations definitely influenced my understanding of the problem.}

\begin{appendix}

\section{Extension of Eq.~(\ref{hcurvedf3}) to discontinuous functions}

If the function $a(x)$ in Eq.~(\ref{hcurvedf3}) does not satisfy conditions
\begin{eqnarray}\label{contcond}
a(0^+) = a(0^-)\,, \quad a(+1) = a(-1)\,,
\end{eqnarray}
\noindent
its derivative is singular at $x=0,\pm 1,$ and the integration by parts used in the transition from Eq.~(\ref{hcurvedf2}) to Eq.~(\ref{hcurvedf3}) is ambiguous. We recall that the functions describing the true flame configuration are actually smooth and periodic, and hence satisfy the conditions (\ref{contcond}), whereas discontinuities arise as the result of simplified description. Therefore, in order to correctly evaluate the integral, one has to turn back to the exact formula (\ref{hcurvedf}) in which all the functions involved are smooth, and apply it to a function $A(x)$ satisfying (\ref{contcond}), whose behavior near the rod or channel walls looks discontinuous from the large-scale point of view. More precisely, $A(x)$ is supposed to vary rapidly for $|x| < R \ll 1$ and near the walls, but normally at the intervals $x_0 < x < 1 - x_0$ and $-1 + x_0 < x < - x_0,$ where it coincides with $a(x).$ Here the positive numbers $R,x_0$ are such that $R < x_0\ll 1$; they have the same meaning as in Sec.~\ref{strategy}. Thus, $$\lim\limits_{R\to 0}A(x) = a(x)\,.$$ ($R \to 0$ implies that $x_0$ also goes to zero.)
Neglecting the anchor dimensions means that the action of $\hat{\EuScript{H}}$ on $a'$ is defined as
\begin{eqnarray}\label{limdef}
\left(\hat{\EuScript{H}}a'\right)(x) = \lim\limits_{R\to 0}\left\{\left(\hat{\EuScript{H}}A'\right)\right\}(x)\,.
\end{eqnarray}
\noindent
To find out how $\hat{\EuScript{H}}$ acts on the derivative of $A(x),$ we replace $a$ by $A$ in Eq.~(\ref{hcurvedf2}), and integrate the right hand side by parts
\begin{eqnarray}\label{hcurvedfa1}
\left(\hat{\EuScript{H}}A'\right)(x) &=& \frac{1 + i
f'(x)}{2}~\fint\limits_{-1}^{+1}
d\eta~A'(\eta)\cot\left\{\frac{\pi}{2}(\eta - x +
i[f(\eta) - f(x)])\right\} \nonumber\\ &=& \frac{1}{2}\frac{d}{dx}\fint\limits_{-1}^{+1}
d\eta~[1 + if'(\eta)]A(\eta)\cot\left\{\frac{\pi}{2}(\eta - x +
i[f(\eta) - f(x)])\right\}\,.
\end{eqnarray}
\noindent The boundary terms vanish here because the integral kernel is $2$-periodic, and $A(x)$ satisfies $A(-1)=A(+1),$ by the assumption. The function $a(x)$ is allowed to have only a finite jump, and so is the slope, $f',$ of the limiting form of the front. Therefore, the last integral in Eq.~(\ref{hcurvedfa1}), in which all  functions are replaced by their limiting expressions, is well-defined, representing a continuously differentiable function for all $|x|\in (0,1).$  Thus, we can write
$$\lim\limits_{R\to 0}\left\{\left(\hat{\EuScript{H}}A'\right)\right\}(x) = \frac{1}{2}\frac{d}{dx}\fint\limits_{-1}^{+1}
d\eta~[1 + if'(\eta)]a(\eta)\cot\left\{\frac{\pi}{2}(\eta - x +
i[f(\eta) - f(x)])\right\}\,,$$ it being understood that $f$ in the integrand is used in its limiting form.

Next, we go over to the large-slope limit. The right hand side of the last equation can be evaluated in this case in exactly the same way as we arrived to Eq.~(\ref{hcurvedf2}). Comparison with Eq.~(\ref{intred}) shows that the role of the function $a(\eta)$ in this equation is now played by $[1 + if'(\eta)] a(\eta),$ the only difference being that the large factor $f'$ comes from the integrand, rather than from the pre-integral factor in Eq.~(\ref{hcurvedf}). Taking this into account, we readily find
\begin{eqnarray}&&
 \left(\hat{\EuScript{H}}a'\right)(x) = \frac{1}{2}\frac{d}{dx}\left[\int\limits_{0}^{1}d\eta \left\{a(\eta)[f'(\eta) - i] + a(-\eta)[f'(-\eta) - i]\right\}\chi(\eta - |x|) \right.\nonumber\\&& \left.- ia(-x)(2|x| - 1)\phantom{\int}\hspace{-0,4cm}\right] = - f'(|x|)\chi(x)\left\{a(|x|) - a(-|x|)\right\} + i\chi(x)\left\{a(|x|) + a(-|x|)\right\} \nonumber\\&& - 2ia(-x)\chi(x) + ia'(-x)(2|x| - 1)\,. \nonumber
\end{eqnarray}
\noindent Using the obvious identities $f'(|x|)\chi(x) = f'(x),$ $\chi(x)\{a(|x|) + a(-|x|) - 2a(-x) \} = a(|x|) - a(-|x|),$ we finally obtain
\begin{eqnarray}&&
 \left(\hat{\EuScript{H}}a'\right)(x) =
 (f'(x) - i)\left\{a(-|x|) - a(|x|)\right\} + ia'(-x)(2|x| - 1)\,, \nonumber
\end{eqnarray}
\noindent which is exactly Eq.~(\ref{hcurvedf3}), as was to be proved. We also observe that the result is independent of the particular choice of $A(x).$

Moreover, it turns out that this formula is valid not only on the open intervals $x\in (-1,0)\cup (0,1),$ but in the whole channel domain $x \in [-1,+1],$ if the derivatives of functions discontinuous at $x=0$ are understood in the sense of distributions. Having in mind possible future applications, let us prove this fact. Note first of all, that if $a(x)$ is discontinuous at $x=0,$ {\it i.e.,} $a(0^+) - a(0^-) \equiv [a]_0 \ne 0,$ then for the function $b(x) = a(x) - [a]_0\chi(x)/2,$ one has $[b]_0 = 0,$ so that Eq.~(\ref{hcurvedf3}) is valid for $b(x).$ Writing $a(x) = b(x) + [a]_0\chi(x)/2,$ we see that since $\hat{\EuScript{H}}$ is a linear operator, it is sufficient to prove the above statement only for the sign function. Let $X(x)$ be its smooth approximation. Take a test function $\phi(x),$ {\it i.e.,} a smooth function that slowly varies for $x\sim R,$ and integrate it with Eq.~(\ref{hcurvedfa1}) over interval $-\Delta \leqslant x \leqslant + \Delta,$ where $\Delta$ is such that $R \ll \Delta < 1.$ We get
\begin{eqnarray}&&
\int\limits_{-\Delta}^{\Delta}dx\phi(x)\left(\hat{\EuScript{H}}X'\right)(x) =  \frac{1}{2}\fint\limits_{-1}^{+1}
d\eta~[1 + if'(\eta)]X(\eta)\nonumber\\&&\times\left[\phi(\Delta)\cot\left\{\frac{\pi}{2}(\eta - \Delta +
i[f(\eta) - f(\Delta)])\right\} - \phi(-\Delta)\cot\left\{\frac{\pi}{2}(\eta + \Delta +
i[f(\eta) - f(\Delta)])\right\}\right] \nonumber\\&& - \frac{1}{2}\int\limits_{-\Delta}^{\Delta}dx\phi'(x)\fint\limits_{-1}^{+1}
d\eta~[1 + if'(\eta)]X(\eta)\cot\left\{\frac{\pi}{2}(\eta - x +
i[f(\eta) - f(x)])\right\}\,.\nonumber
\end{eqnarray}
\noindent All integrals on the right are well-defined in the limit $R \to 0,$ so that $X(\eta)$ can be replaced by $\chi(\eta).$ Then the $\eta$-integrations are readily done because the primitives are $\ln\sin\{\cdot\}$. For example,
\begin{eqnarray}&&
\fint\limits_{-1}^{+1} d\eta~[1 + if'(\eta)]X(\eta)\cot\left\{\frac{\pi}{2}(\eta - \Delta + i[f(\eta) - f(\Delta)])\right\} \nonumber\\&& = \frac{2}{\pi}\left[\fint\limits_{0}^{+1} - \int\limits_{-1}^{0}\right]d\ln\sin\left\{\frac{\pi}{2}(\eta - \Delta + i[f(\eta) - f(\Delta)])\right\} = 2[f(1) - i] - 4[f(\Delta) -i\Delta ]\,,\nonumber
\end{eqnarray}
\noindent where it is taken into account that the front slope is large for $|\eta|\gg R.$ A simple calculation gives
$$\int\limits_{-\Delta}^{\Delta}dx\phi(x)\left(\hat{\EuScript{H}}X'\right)(x) = -2i\phi(0) - 2\int\limits_{-\Delta}^{\Delta}dx\phi(x)[f'(x) - i]\,.$$
Finally, since $\phi(x)$ is independent of $R,$ the limit of this equation for $R \to 0$ can be written using the definitions of Dirac $\delta$-function and (\ref{limdef}) as
$$\int\limits_{-\Delta}^{\Delta}dx\phi(x)\left(\hat{\EuScript{H}}\chi'\right)(x) = -2i\int\limits_{-\Delta}^{\Delta}dx\delta(x)\phi(x) - 2\int\limits_{-\Delta}^{\Delta}dx\phi(x)[f'(x) - i]\,,$$
which in view of arbitrariness of $\phi(x)$ yields
$$\left(\hat{\EuScript{H}}\chi'\right)(x) = -2i\delta(x) - 2[f'(x) - i]\,.$$
By virtue of the relations $\chi'(x) = 2\delta(x),$ $|x|\delta(x) =0,$ understood in the sense of distributions, this is just Eq.~(\ref{hcurvedf3}) for $a=\chi.$

\end{appendix}

~\newpage

\bibliography{references}

~\newpage
~\\\\
{\large\bf List of figures}\\\\ Schematics of a flame anchored in a channel.
Shown is the right half of the channel, in the case of anchor placed in its middle.
As is, this figure also represents the flame anchored at the channel wall.
\dotfill 17\\ Schematics of the flow structure near a circular rod in the case $w_0 > 0$ and moderate $U.$ A stream line crossing the flame front in the matching region is shown. For larger $U,$ the point $x_0$ is closer to the rod. \dotfill 18\\
Near-the-rod behavior of $Q(x)$ according to the true local solution (solid line), and  extrapolated large-scale solution (broken line)\dotfill 19\\
Numerical solution for the flame front position, longitudinal ($u$) and transversal ($w$) fresh gas velocity on-shell in the case $\theta = 8,$ $U = 60,$ $w_0 = 0.2.$ \dotfill 20\\
Same for $\theta = 6,$ $U = 40,$ $w_0 = 2.$ \dotfill 21\\
Same for $\theta = 3,$ $U = 20,$ $w_0 = -0.1.$ \dotfill 22\\

~\newpage
~\newpage
~\newpage

~\newpage

~\newpage

~\newpage

\end{document}